# Least-false Cox coefficients under affine follow-up contamination: exact continuous- and grouped-time benchmarks

**Isfandiyor Akhmedov**
School of Banking and Finance
Management Development Institute
of Singapore in Tashkent, Tashkent, Uzbekistan
Corresponding author: iakhmedov@mdist.uz
ORCID: https://orcid.org/0009-0008-4233-4291

**Abstract**

Covariates summarized over a subject's completed follow-up are sometimes entered into Cox regression as though observed at baseline. This practice incorporates future event or censoring information and changes both the estimand and its sampling behavior. We analyze an affine class in which a genuine baseline covariate is contaminated by realized follow-up time. Treating partial likelihood as an observed-data estimation criterion, we derive the population score and characterize its unique least-false coefficient. Exact continuous-time benchmarks show scale reduction and saturation under strong contamination, while administrative censoring destroys the reduction and may produce overshoot. Grouping exit times with Breslow ties changes the geometry: the coefficient has a single hump and eventually returns to zero even though the induced association diverges. We also derive an observed-data influence function and show why model-based variance can be either too small or too large. A subject-level sandwich consistently estimates uncertainty around the least-false target under the stated conditions, but it does not correct the target itself.

## Introduction

A variable measured or summarised over a subject's completed follow-up is sometimes entered into a Cox model as though it had been known at baseline. The practice is easy to fall into because the resulting variable looks like an ordinary covariate: one number per subject, available at the time of analysis, with no time index attached. Three examples recur.

*Attained age at exit.* A study records age at entry and age at the end of observation, and the second is used as "age". *Average age over observed follow-up.* The analyst averages entry and exit age, obtaining $A + 1/2\,(T \wedge C)$ for a subject entering at age $A$. *Longitudinal summaries.* A biomarker is measured repeatedly, and its mean, area under the curve, slope or maximum over each subject's available trajectory is entered as a baseline predictor.

In each case the length of the trajectory being summarised is $T \wedge C$. The summary therefore contains outcome information before it is placed in the model. The consequence is not subtle: a subject who fails early has accumulated less of whatever is being summed, so small values of the constructed covariate travel with early failure. The constructed covariate can therefore acquire a *protective* appearance. Under the null benchmark studied below, in which the underlying baseline covariate has no effect at all, the population score cannot point in the opposite direction, and is strictly protective whenever the analysis contains an informative event before the horizon.

## Motivation and related work

The logical error was identified and named long ago. Wolfe and Strawderman (1996) warn that a covariate averaged over a patient's entire follow-up should almost never be used as a baseline covariate, recommend instead the baseline value or a cumulative average updated to the current time, and show that the practice can reverse the apparent direction of an association. They also draw the distinction that matters here, between information that merely became available late and information that was actually generated in the future. The broader methodological literature on time-dependent covariates makes the same point in general terms: Fisher and Lin (1999) emphasise that the interrelationship between outcome and covariate over time can generate bias unless it is well understood, and that internal covariates require particular care. We do not claim the phenomenon itself as new.

That the problem is live rather than historical is visible in applied work on longitudinal summaries. Dodd et al. (2014) study biomarker trajectories in which non-survivors are observed for shorter periods precisely because they die, explain why direct comparisons of derived summaries across unequal observation lengths are inappropriate, and propose matching subjects over a common interval. Their setting is not ours — they compare groups rather than analyse a partial likelihood — but it documents the same construction appearing in later applied work.

On the theoretical side, the behaviour of the Cox estimator under misspecification is well developed. Struthers and Kalbfleisch (1986) show that the maximum partial-likelihood estimator converges to an implicitly defined least-false parameter; Gerds and Schumacher (2001) treat functional misspecification of a covariate; Lin and Wei (1989) give robust inference for the misspecified model. These are the right tools in spirit, and Section 6 uses their language. But their standard representations are derived for admissible fitted covariates, and they do not directly cover a covariate that contains $T \wedge C$: once $W$ is constructed, it is no longer predictable in the observed filtration, and censoring is no longer conditionally independent of failure given the fitted covariate. Adjacent methodology on error-prone and time-varying exposures, such as the risk-set calibration of Liao et al. (2011), addresses a genuinely different problem: a surrogate for a latent time-varying exposure observed with measurement error, rather than a covariate constructed deterministically from the realised event-or-censoring time.

### The gap

What appears to be missing is an exact population theory for the Cox coefficient obtained when a fixed fitted covariate contains an affine component of realised follow-up,

$$W = A + \gamma\,(T \wedge C).$$

The warning literature establishes that this is a mistake and that it can flip a sign. It does not answer what the mistake costs. Specifically:

1. What is the exact least-false coefficient that the naive analysis targets?
2. How does it depend on contamination strength, on the censoring law, and on the distribution of the honest covariate $A$?
3. What changes when exit times are reported on a grid and the analysis must handle ties?

4. Why does the default Cox variance sometimes understate and sometimes overstate the uncertainty?

5. Can the robust sandwich remain valid when the martingale argument that ordinarily justifies it does not apply?

The affine class is worth isolating because $\gamma$ is not a nuisance parameter to be estimated: it is fixed by the construction of the indicator, with $\gamma = 1$ for age at exit and $\gamma = 1/2$ for average age during follow-up. This removes one unknown from the problem and makes exact answers available.

## Contributions

**An observed-data target.** The generic ingredients here are standard: a misspecified Cox fit converges to an implicitly defined least-false parameter (Struthers and Kalbfleisch 1986; Gerds and Schumacher 2001), and plug-in $M$-estimation limits are routine. What is not standard is that they apply at all, since $W$ is inadmissible as a fitted covariate; we therefore derive the population score directly from the joint law of $(U, \Delta, W)$, without a compensator representation. This is an enabling reduction rather than a new general theorem about misspecified Cox regression. Under the null benchmark assumptions, the score at the origin is non-positive for every $\gamma > 0$, and strictly negative under an explicit informative-event condition, for arbitrary marginal laws satisfying the stated regularity conditions: the spurious protective association is a property of the construction rather than of any particular distributional choice. We establish general concavity, score-crossing existence and consistency results, and obtain an exact necessary-and-sufficient existence criterion in the memoryless benchmark.

**Exact coefficient geometry.** In the Gaussian memoryless benchmark the coefficient has a closed form and the whole problem reduces to one dimensionless ratio; the coefficient saturates rather than diverging, at a level that depends on the shape of the law of $A$ and not only on its variance. Administrative censoring breaks the reduction. On a reporting grid the exact Breslow equation is two-dimensional, the raw coefficient has a single Lambert-$W$ hump and returns to zero, the limits in grid width and contamination strength do not commute, and the standardised pseudo-effect diverges logarithmically rather than linearly.

**Inference.** The sandwich principle, the generic possibility that the information equality fails, and the algebra of Cox score residuals are all long established (Lin and Wei 1989); our influence-function derivation justifies them for a data construction their standard proofs do not cover, and the substantive new content is the exact evaluation. We obtain the observed-data influence function, exact expressions for the bread $J$ and the meat $K$, and a proof that the information equality fails for every non-zero contamination in the uncensored benchmark. Under exponential censoring the direction of failure is governed by an explicit boundary $p_*(\rho)$ in the plane of event fraction and contamination ratio; a matched counterexample shows that the boundary does not extend beyond the competing-exponential family.

**Repair and its limits.** At any fixed analysis horizon, the subject-level sandwich consistently estimates the sampling variance around the least-false target on both sides of the boundary. For the unrestricted competing-exponential Cox fit, the corresponding statement is conditional on the $L^2$ tail conditions stated in Supplementary Material. It does not repair the target. In the affine class the admissible baseline component is recovered exactly as $A = W - \gamma U$, so the correction belongs at the stage of covariate construction rather than variance estimation.

The finite-horizon asymptotic theory is unconditional under the stated regularity conditions. Removal of the horizon for the competing-exponential central limit theorem is conditional on the $L^2$ tail conditions stated in Supplementary Material.

**Organisation**

Section 2 fixes the model and notation. Section 3 constructs the observed-data target and its asymptotics. Sections 4 and 5 study the continuous-time and grouped-time geometry of the least-false coefficient. Section 6 develops inference. Section 7 discusses interpretation and limitations, and Supplementary Material contains the technical proofs.

# The model

We work throughout with the proportional-hazards model of Cox (1972) and its partial likelihood. For subjects $i = 1, \dots, n$, independent and identically distributed: $A_i$ is a baseline covariate with $\mathbb{E}A = 0$ and $\text{Var}A = \sigma_A^2$; $T_i$ is the event time, with hazard $\lambda(t \mid A) = \lambda_0(t)e^{\beta A}$; $C_i$ is the censoring time; and

$$U_i = T_i \wedge C_i, \qquad \Delta_i = \mathbf{1}\{T_i \le C_i\}, \qquad W_i = A_i + \gamma U_i, \qquad \mathcal{O}_i = (U_i, \Delta_i, W_i).$$

Here $W$ is the *contaminated covariate actually fitted*, with $\gamma$ known by construction; $\theta$ is the coefficient the analyst estimates for $W$; and $\tau$ is an analysis horizon with $\mathbb{P}(U \ge \tau) > 0$.

Throughout Sections 3–5, and in the exact inferential benchmarks of Section 6, we work under the *null specification*, in which we impose

$$\beta = 0, \qquad A \perp (T, C), \qquad T \perp C. \tag{1}$$

These are three separate assumptions and all three are used. Setting $\beta = 0$ makes the law of $T$ free of $A$, but by itself gives neither $A \perp C$ nor $T \perp C$; the factorisations $S_U = S_T S_C$ and $e^{\theta W} = e^{\theta A} e^{\theta \gamma U}$ below require the other two. The benchmark is the case in which the covariate carries no effect at all, so that every association the analyst recovers is contamination. Write $S_T, S_C$ for the survival functions of $T, C$, $S_U = S_T S_C$, and

$$F_1(\mathrm{d}t) = \mathbb{P}(U \in \mathrm{d}t,\ \Delta = 1) = f_T(t) S_C(t)\,\mathrm{d}t.$$

Let $M_A(\theta) = \mathbb{E}e^{\theta A}$, $K_A = \log M_A$, and let $K_A{}'$ denote its derivative.

# Observed-data target and asymptotic foundation

## The observed-data criterion

The analyst fits a Cox model with the single covariate $W$. Discarding an additive constant, the normalised partial log-likelihood over $[0, \tau]$ is

$$\mathbb{M}_{n,\tau}(\theta) = \frac{1}{n}\sum_{i=1}^{n} \Delta_i\, \mathbf{1}\{U_i \le \tau\}\Big[\theta W_i - \log \mathbb{S}_n^{(0)}(\theta, U_i)\Big], \qquad \mathbb{S}_n^{(k)}(\theta, t) = \frac{1}{n}\sum_{j=1}^{n} \mathbf{1}\{U_j \ge t\} W_j^k e^{\theta W_j},$$

with score

$$\Psi_{n,\tau}(\theta) = \frac{1}{n}\sum_{i=1}^{n} \Delta_i\, \mathbf{1}\{U_i \le \tau\} \left[ W_i - \frac{\mathbb{S}_n^{(1)}(\theta, U_i)}{\mathbb{S}_n^{(0)}(\theta, U_i)} \right].$$

We stress at the outset what this object is and is not. It is a random criterion computed from $n$ independent copies of $O = (U, \Delta, W)$, and we study it purely as such, as a misspecified $M$-estimation problem. It is *not* being used as a partial likelihood for a correctly specified proportional-hazards model, and no step below asserts that the conditional hazard of $T$ given $W$ is of Cox form.

**Why the standard argument does not transfer**

The classical treatment of misspecified Cox models (Struthers and Kalbfleisch 1986; Lin and Wei 1989; Gerds and Schumacher 2001) derives the limiting equation for $\hat{\theta}$ from a compensator representation of the score, and that derivation is carried out under a censoring assumption stated relative to the fitted covariate: censoring is conditionally independent of the failure time given the covariate in the model. Under our construction that assumption fails, and it fails structurally rather than by accident of parametrisation.

**Lemma 3.1 (Failure of conditional independence).** Let $A \perp (T, C)$ with $T \perp C$, let $T$ and $C$ have positive continuous densities on a common open interval support, let $A$ have a positive continuous density $f_A$, write $\ell = \log f_A$, and let $\gamma \neq 0$, $W = A + \gamma(T \wedge C)$. Suppose that for $w$ in a set of positive measure the map $u \mapsto \ell(w - \gamma u)$ is non-constant on the support of $T \wedge C$. Then $T \not\perp C \mid W$.

*Proof.* On the event $\{W = w\}$ the joint conditional density of $(T, C)$ is proportional to $f_T(t) f_C(c) f_A\big(w - \gamma \min(t, c)\big)$. Conditional independence would require this to factorise as $g(t)h(c)$, equivalently that $\psi(t, c) = \ell\big(w - \gamma \min(t, c)\big)$ be additively separable, and additive separability is equivalent to the vanishing of every mixed second difference.

Fix $w$ in the assumed set. Because $u \mapsto \ell(w - \gamma u)$ is non-constant on the support, there exist two points in the support at which it takes *different values*; label them $c_1 < t_2$, so that $\ell(w - \gamma t_2) \neq \ell(w - \gamma c_1)$. This is the step at which one must be careful: distinct arguments alone do not give distinct values, since $\ell$ may be symmetric, as it is in the Gaussian case, so the pair must be selected from non-constancy rather than assumed generic. Now pick $t_1 < c_1$ and $c_2 > t_2$ in the support, which is possible because the support is an open interval. With $t_1 < c_1 < t_2 < c_2$ one has $\min(t_2, c_2) = t_2$, $\min(t_2, c_1) = c_1$ and $\min(t_1, c_2) = \min(t_1, c_1) = t_1$, so

$$\psi(t_2, c_2) - \psi(t_2, c_1) - \psi(t_1, c_2) + \psi(t_1, c_1) = \ell(w - \gamma t_2) - \ell(w - \gamma c_1) \neq 0.$$

By continuity the mixed difference is non-zero on a neighbourhood, hence on a set of positive measure, and no separable representation exists.

The hypothesis is mild: it holds whenever $f_A$ is non-constant on every non-degenerate interval, which covers every strictly log-concave density and in particular the Gaussian benchmark, where $\ell(x) = -x^2/2\sigma_A^2 + \text{const}$.

**Remark (Where the interaction comes from).** Restricted to the half-plane $\{t < c\}$ one has $\psi(t, c) = \ell(w - \gamma t)$, which *is* a function of $t$ alone; likewise on $\{c < t\}$. The obstruction is exactly the switch between the two branches: knowing $W$ and knowing that $T$ is large forces $C$ to be small, because otherwise $A$ would have to take an implausible value. Censoring and failure become competitors for the same budget $W - A$.

**Remark ($W$ is not a baseline covariate).** The value of $W$ is not determined until $T \wedge C$ has been realised, so $W$ is not $\mathcal{F}_0$-measurable and the process $t \mapsto W$ is not predictable with respect to the observed filtration. The martingale machinery that produces the usual compensator identity therefore has no purchase here even before the censoring condition is considered.

Consequently the population equation defining the limit of $\hat{\theta}$ must be derived from the joint law of $(U, \Delta, W)$ directly. What is unavailable is the *representation* of the limit, not the existence of a limit: the estimator is a $Z$-estimator of an i.i.d. empirical criterion, and its convergence (Section 3.5) requires neither predictability nor a censoring assumption.

## Population score and residual-life representation

Define, for $k = 0,1,2$,

$$s^{(k)}(\theta, t) = \mathbb{E}\big[\mathbf{1}\{U \geq t\} W^k e^{\theta W}\big], \qquad \mu_\theta(t) = \frac{s^{(1)}(\theta, t)}{s^{(0)}(\theta, t)}.$$

Thus $\mu_\theta(t)$ is the mean of $W$ in the risk set at $t$ under the exponential tilt $e^{\theta W}$, a purely observational quantity, defined whether or not any model holds.

**Proposition 3.4 (Observed-data score).** Suppose $\mathbb{E}\big[(1 + |W|)e^{\theta W}\big] < \infty$ and $\inf_{t \leq \tau} s^{(0)}(\theta, t) > 0$. Then $\Psi_{n,\tau}(\theta) \to \Psi_\tau(\theta)$ in probability, where

$$\Psi_\tau(\theta) = \mathbb{E}\big[\Delta \mathbf{1}\{U \leq \tau\}\big(W - \mu_\theta(U)\big)\big] = \int_{[0,\tau]} [\mathbb{E}(W \mid U = t, \Delta = 1) - \mu_\theta(t)]\; F_1(\mathrm{d}t).$$

This is a statement about the law of $(U, \Delta, W)$ and nothing else. It does not decompose $U$ into failure and censoring components, does not require $T \perp C \mid W$, and does not require the conditional hazard given $W$ to be proportional. The least-false parameter is *defined* as the root of this equation, $\Psi_\tau\big(\theta_\tau^\dagger\big) = 0$, with existence and uniqueness treated in Section 3.4.

### The residual-life form

For the affine construction the two terms can be evaluated explicitly, and doing so produces every closed form in Sections 4 and 5. Under the null-benchmark assumptions (1), with $W = A + \gamma U$:

- conditionally on $\{U = t, \Delta = 1\}$ we have $W = A + \gamma t$ with $A$ unaffected by the conditioning, so $\mathbb{E}(W \mid U = t, \Delta = 1) = \gamma t$;
- the tilt factorises, $e^{\theta W} = e^{\theta A} e^{\theta \gamma U}$ with independent factors, so

  $$s^{(0)}(\theta, t) = M_A(\theta)\, \mathbb{E}\big[\mathbf{1}\{U \geq t\} e^{\theta \gamma U}\big], \qquad \mu_\theta(t) = K_A{}'(\theta) + \gamma\, \mathbb{E}_\theta(U \mid U \geq t),$$

  where $\mathbb{E}_\theta$ denotes expectation under the tilt proportional to $e^{\theta \gamma u}$.

Writing $R_\theta(t) = \mathbb{E}_\theta(U - t \mid U \geq t)$ for the *tilted mean residual life* at $t$:

**Corollary 3.5 (Residual-life representation).**

$$\Psi_\tau(\theta) = -K_A{}'(\theta)\, F_1([0, \tau]) \;-\; \gamma \int_{[0,\tau]} R_\theta\,(t)\, F_1(\mathrm{d}t). \qquad (2)$$

The two terms are the two mechanisms. The first is the price of tilting the baseline covariate; the second is the contamination proper, and it is an average of residual lifetimes, the quantity that a subject failing at $t$ *did not* accumulate relative to those still at risk. Three consequences follow immediately.

**(i) The sign at the origin is distribution-free.**

Since $K_A{}'(0) = \mathbb{E}A = 0$,

$$\Psi_\tau(0) = -\gamma \int_{[0,\tau]} R_0\,(t)\,F_1(\mathrm{d}t) \le 0 \qquad \text{for every } \gamma > 0,$$

whatever the laws of $A$, $T$ and $C$, with

$$\Psi_\tau(0) < 0 \qquad \Leftrightarrow \qquad \int_{[0,\tau]} R_0\,(t)\,F_1(\mathrm{d}t) > 0. \tag{3}$$

The integrand $R_0(t) = \mathbb{E}(U - t \mid U \ge t)$ is non-negative; it is strictly positive precisely at those risk-set times for which positive residual follow-up remains with positive conditional probability. Hence (3) says exactly that the event subdistribution must place positive mass on such times: the design must contain an *informative event* before $\tau$. The condition is not vacuous. Take $A \sim N(0,1)$, $T = 2 + E$ with $E \sim \mathrm{Exp}(1)$, $C \equiv 10$ and $\tau = 1$: every independence assumption holds and $\mathbb{P}(U \ge \tau) = 1$, but no event can occur before $\tau$, so $F_1([0,\tau]) = 0$, the criterion is identically zero and $\Psi_\tau(0) = 0$. A convenient sufficient condition for continuous event times is $F_1([0,\tau)) > 0$ together with $\mathbb{P}(U \ge \tau) > 0$. Subject to it, a spurious *protective* association is not a feature of any particular model but of the construction itself.

**(ii) The memoryless benchmark solves in closed form, and the horizon drops out.**

If $U \sim \mathrm{Exp}(\kappa)$ — in particular when $T$ and $C$ are independent exponentials with $\kappa = \lambda_T + \lambda_C$, but more generally whenever the hazards satisfy $\lambda_T(t) + \lambda_C(t) \equiv \kappa$, so that the individual hazards may vary with time — the tilted law of $U - t$ given $U \ge t$ is $\mathrm{Exp}(\kappa - \theta\gamma)$ for every $t$, so $R_\theta(t) \equiv (\kappa - \theta\gamma)^{-1}$ and

$$\Psi_\tau(\theta) = -F_1([0,\tau])\,b(\theta), \qquad b(\theta) = K_A{}'(\theta) + \frac{\gamma}{\kappa - \theta\gamma}. \tag{4}$$

The horizon enters only through the positive factor $F_1([0,\tau])$: in this benchmark $\theta_\tau^\dagger = \theta^\dagger$ for every $\tau$. For Gaussian $A$, $K_A{}'(\theta) = \sigma_A^2\theta$ and

$$\theta^\dagger = \frac{\kappa - \sqrt{\kappa^2 + 4\gamma^2/\sigma_A^2}}{2\gamma},$$

the closed form used throughout Section 4.

**(iii) The large-contamination limit is a statement about $A$ alone.**

As $\gamma \to \infty$, $\gamma/(\kappa - \theta\gamma) \to -1/\theta$, so (4) collapses to $\theta K_A{}'(\theta) = 1$, free of $\kappa$ and hence, *within the memoryless benchmark*, of the baseline hazard and the censoring distribution. Existence of a finite limit requires $\sup_{\theta<0}\theta K_A{}'(\theta) > 1$, a condition on the tail of $A$ and not a regularity technicality; see Section 3.4.

**Remark 3.6 (Numerical verification).** Corollary 3.5 was checked against the empirical score outside the memoryless case. Roots of $\Psi$ obtained by quadrature from (2), against Monte Carlo roots of the empirical Cox score at $n = 8 \times 10^5$, with $A$ standard Gaussian, $T$ Weibull with shape $k$ and $C$ exponential with rate $\lambda_C$:

| $k$ | $\lambda_C$ | $\gamma$ | root of $\Psi$ | Monte Carlo | sign$\Psi(0)$ |
|---|---|---|---|---|---|
| 1.0 | $10^{-4}$ | 1.0 | −0.61801 | −0.61888 | − |
| 1.5 | 0.5 | 1.0 | −0.42424 | −0.42440 | − |
| 2.0 | 0.3 | 2.0 | −0.61044 | −0.61130 | − |
| 0.7 | 0.8 | 1.5 | −0.57992 | −0.58024 | − |

Agreement is to Monte Carlo accuracy in all cases, including decreasing hazard ($k = 0.7$) and increasing hazard ($k = 2.0$), and $\Psi(0) < 0$ throughout, as (i) requires.

## Existence, uniqueness, and consistency

Define the population criterion as the limit of $\mathbb{M}_{n,\tau}$,

$$\begin{aligned}\mathbb{M}_\tau(\theta) &= \mathbb{E}\big[\Delta \mathbf{1}\{U \le \tau\}\{\theta W - \log s^{(0)}(\theta, U)\}\big] \\ &= \int_{[0,\tau]} \big[\theta\, \mathbb{E}(W \mid U = t, \Delta = 1) - \log s^{(0)}(\theta, t)\big]\, F_1(\mathrm{d}t),\end{aligned}$$

so that $\mathbb{M}_\tau{}' = \Psi_\tau$ by Proposition 3.4. Write $\Theta_{\max} = \mathrm{int}\{\theta\colon \mathbb{E}\big[(1 + W^2)e^{\theta W}\big] < \infty\}$ and, for $\theta \in \Theta_{\max}$,

$$v_\theta(t) = \frac{s^{(2)}(\theta, t)}{s^{(0)}(\theta, t)} - \Big(\frac{s^{(1)}(\theta, t)}{s^{(0)}(\theta, t)}\Big)^2,$$

the variance of $W$ under the measure proportional to $\mathbf{1}\{U \ge t\}e^{\theta W}\,\mathrm{d}\mathbb{P}$, that is, the tilted risk-set variance.

**Proposition 3.7 (Concavity; at most one finite maximiser).** On $\Theta_{\max}$,

$$\mathbb{M}_\tau{}''(\theta) = -\int_{[0,\tau]} v_\theta\,(t)\, F_1(\mathrm{d}t) \;\le\; 0.$$

If $v_\theta(t) > 0$ on a set of $t$ of positive $F_1$-measure, then $\mathbb{M}_\tau$ is strictly concave, $\Psi_\tau$ is strictly decreasing, and $\mathbb{M}_\tau$ has at most one finite maximiser.

This holds *for any joint law of* $(U, \Delta, W)$: a variance is non-negative whatever generated the data, so no model needs to be correct for $\mathbb{M}_\tau$ to be concave. Uniqueness of $\theta_\tau^\dagger$ therefore need not be assumed; and the non-degeneracy requirement is mild, failing only if $W$ is almost surely constant within risk sets. In the affine construction the risk set at any $t$ with $s^{(0)}(\theta, t) > 0$ contains the non-degenerate variable $A$, so $v_\theta(t) > 0$ throughout.

Existence is the substantive condition and must be stated separately.

**Proposition 3.8 (Existence by score crossing).** Suppose there exist $\theta_- < \theta_+$ in $\Theta_{\max}$ with $\Psi_\tau(\theta_-) > 0 > \Psi_\tau(\theta_+)$. Then, under the strict concavity of Proposition 3.7, $\mathbb{M}_\tau$ has a unique finite maximiser $\theta_\tau^\dagger \in (\theta_-, \theta_+)$.

In the memoryless benchmark the crossing condition admits an exact, checkable form. By (4),

$$b'(\theta) = K_A''(\theta) + \frac{\gamma^2}{(\kappa - \theta\gamma)^2} > 0$$

for non-degenerate $A$, since $K_A''$ is a variance; so $b$ is strictly increasing and $b(0) = \gamma/\kappa > 0$.

**Corollary 3.9 (Exact existence criterion, memoryless benchmark).** Let $U \sim \mathrm{Exp}(\kappa), \gamma > 0, A$ non-degenerate with $\mathbb{E}A = 0$, and let $\theta_L \in [-\infty, 0)$ be the left endpoint of the domain of $K_A$. Then a finite interior root $\theta^\dagger$ exists if and only if

$$\lim_{\theta \downarrow \theta_L} b(\theta) < 0.$$

In particular, if $\theta_L = -\infty$ then $K_A'(\theta) \to \operatorname{ess\,inf} A$ while $\gamma/(\kappa - \theta\gamma) \downarrow 0$, so the limit equals $\operatorname{ess\,inf} A$, which is strictly negative for any non-degenerate mean-zero $A$; existence is then automatic.

The "if and only if" is available precisely because $b$ is strictly monotone, so a single boundary value decides the question. When $\theta_L$ is finite the criterion must be checked: a finite left endpoint with $K_A'$ bounded there makes failure *possible*, but the root still exists whenever the boundary value of $b$ is negative. The criterion is therefore a genuine restriction on heavy left tails of $A$ rather than a regularity technicality. It is related to, but not the same as, the requirement $\sup_{\theta<0} \theta K_A'(\theta) > 1$ governing the large-contamination limit: the latter concerns existence of $\lim_{\gamma\to\infty} \theta^\dagger$, the former existence at each fixed $\gamma$. Both are conditions on the left tail of $A$, and both can fail only when $\theta_L$ is finite, but neither implies the other.

**Remark (Three existence questions must be kept apart).**

1. *Existence of the population root* $\theta_\tau^\dagger$: Proposition 3.8, a property of the law.
2. *Existence of a finite empirical maximiser* at a given $n$: a property of the realised sample.
3. *Finite-sample separation*: with positive probability the ordering of $(W_i, U_i, \Delta_i)$ is such that $\Psi_{n,\tau}$ has no zero and $\hat{\theta}_{n,\tau} = -\infty$. As $\gamma$ grows at fixed $n$ this becomes the typical outcome *provided the sample contains at least one event occurring while another subject is still at risk*, since such a failing subject then holds the smallest $W$ in its risk set. Monotone likelihood and divergent Cox estimates are a familiar phenomenon in their own right, and penalised remedies for them are well established (Heinze and Schemper 2001); what the affine construction contributes is an explicit mechanism driving them.

Separation at fixed $n$ does not contradict consistency: when the population root is finite and interior, the localisation argument of Section 3.6 shows that its probability tends to zero; and at fixed $n$ the statement above is itself conditional on the sample containing an informative event. In the competing-exponential benchmark with event probability $p$, the marks being independent of the exit ranks, $\lim_{\gamma\to\infty} \mathbb{P}(\text{separation}) = 1 - (1-p)^{n-1}$, which is strictly below one at fixed $n$; here too the limits fail to commute, since $\lim_{\gamma\to\infty} \lim_{n\to\infty} \mathbb{P}(\text{separation}) = 0$ whereas $\lim_{n\to\infty} \lim_{\gamma\to\infty} \mathbb{P}(\text{separation}) = 1$. On a fixed grid, ties interrupt the deterministic ordering mechanism that drives separation in continuous time; separation is therefore no longer guaranteed, and whether it occurs depends on the tied-event configuration and on the tie rule. What separation does mean is that *no* statement about $\hat{\theta}$ can be made uniformly over $\gamma$ at fixed $n$: the asymptotics are pointwise in $\gamma$.

**Consistency on a finite horizon**

Fix $\tau$ with $\mathbb{P}(U \geq \tau) > 0$ and assume

$\Theta$ is compact, $\Theta \Subset \Theta_{\max}$, and $\theta_\tau^\dagger \in \mathrm{int}\Theta$;

$\inf_{\theta \in \Theta,\, t \leq \tau} s^{(0)}(\theta, t) > 0$;

$v_\theta(t) > 0$ on a set of positive $F_1$-measure, for every $\theta \in \Theta$.

Condition (C1) is deliberately one-sided and local. The symmetric requirement $\mathbb{E}\big[(1 + |W|^2)e^{K|W|}\big] < \infty$ for $K > \sup_\Theta|\theta|$, which one might write by reflex, is too strong here: with $\gamma > 0$ and $U \sim \mathrm{Exp}(\kappa)$ it forces $K\gamma < \kappa$, hence $\sup_\Theta|\theta| < \kappa/\gamma \to 0$, and so excludes the very regime of interest, in which $\theta^\dagger \to -1/\sigma_A$. No such difficulty arises on the negative half-line: for $\theta \leq 0$ and $\gamma > 0$ we have $e^{\theta\gamma U} \leq 1$, so only left exponential moments of $A$ and ordinary polynomial moments of $U$ are needed.

Condition (C2) is the standard bounded-away-from-zero requirement on the risk-set denominator, familiar from the counting-process theory of Andersen and Gill (1982), and it is the effective-exposure floor here. Under independent censoring $s^{(0)}(0, \tau) = \mathbb{P}(U \geq \tau) = S_T(\tau)S_C(\tau)$, the same quantity that governs non-asymptotic bounds for the Nelson–Aalen and Kaplan–Meier estimators, and the reason a finite horizon is needed at all.

Two estimators must be distinguished. The *constrained* maximiser

$$\hat{\theta}_{n,\tau}^{\Theta} \in \arg\max_{\theta \in \Theta} \mathbb{M}_{n,\tau}(\theta) \qquad \text{(any measurable selection)}$$

always exists, because $\Theta$ is compact and $\mathbb{M}_{n,\tau}$ is continuous; the maximiser need not be unique at finite $n$ — with no events before $\tau$ the criterion is constant — so a selection rather than an equality is the right definition, uniqueness being recovered only in the limit through population strict concavity. It exists even on samples exhibiting separation, where it sits at an endpoint of $\Theta$. The *unrestricted* maximiser $\hat{\theta}_{n,\tau}$ over all of $\mathbb{R}$, the object a Cox routine actually returns, need not exist. Theorem 3.11 is a statement about the first; the second is recovered afterwards.

**Theorem 3.11 (Finite-horizon consistency).** Under (C1)–(C3),

$$\sup_{\theta \in \Theta} |\mathbb{M}_{n,\tau}(\theta) - \mathbb{M}_\tau(\theta)| \xrightarrow{p} 0, \qquad \hat{\theta}_{n,\tau}^{\Theta} \xrightarrow{p} \theta_\tau^\dagger.$$

If moreover $\theta_- < \theta_\tau^\dagger < \theta_+$ lie in $\mathrm{int}\Theta$ with $\Psi_\tau(\theta_-) > 0 > \Psi_\tau(\theta_+)$, then $\mathbb{P}\Big(\Psi_{n,\tau}(\theta_-) > 0 > \Psi_{n,\tau}(\theta_+)\Big) \to 1$, and on that event the unrestricted maximiser exists, is finite and lies in $(\theta_-, \theta_+)$; any measurable selection $\hat{\theta}_{n,\tau}^{\Theta}$ from the constrained maximising set then lies in the same interval, and the two coincide whenever the empirical criterion is strictly concave there. Hence $\hat{\theta}_{n,\tau} \xrightarrow{p} \theta_\tau^\dagger$ as well.

*Proof.* The classes $\{\mathbf{1}\{u \geq t\} w^k e^{\theta w} : \theta \in \Theta,\ t \in [0, \tau]\}$, $k = 0,1,2$, are products of a VC class of indicators with a fixed measurable envelope $(1 + w^2)\big(e^{\theta_- w} + e^{\theta_+ w}\big)$, integrable by (C1); they are therefore Glivenko–Cantelli, and $\sup_{\theta,t}\Big|\mathbb{S}_n^{(k)}(\theta, t) - s^{(k)}(\theta, t)\Big| \to 0$ almost surely.

The criterion is not a plain empirical mean, since its integrand depends on $\mathbb{P}_n$ through $\mathbb{S}_n^{(0)}$. Writing $f_{P,\theta}(o) = \delta \mathbf{1}\{u \le \tau\}\{\theta w - \log s_{P,\theta}^{(0)}(u)\}$, decompose

$$\mathbb{P}_n f_{\mathbb{P}_n,\theta} - P f_{P,\theta} = \mathbb{P}_n\big(f_{\mathbb{P}_n,\theta} - f_{P,\theta}\big) + (\mathbb{P}_n - P) f_{P,\theta}.$$

The second term is uniformly small by the Glivenko–Cantelli property of $\{f_{P,\theta} : \theta \in \Theta\}$. For the first, (C2) bounds $s^{(0)}$ away from zero on $[0,\tau]$, so $x \mapsto \log x$ is Lipschitz on the relevant range and $\sup_\theta \| f_{\mathbb{P}_n,\theta} - f_{P,\theta} \|_\infty \lesssim \sup_{\theta,t}\left|\mathbb{S}_n^{(0)} - s^{(0)}\right| \to 0$ almost surely, while $\mathbb{P}_n$ is a probability measure. Hence the criteria converge uniformly. Strict concavity (C3) makes $\theta_\tau^\dagger$ a well-separated maximiser, and the argmax theorem (Vaart 1998, Thm.5.7) applies to any measurable selection from the maximising set.

Every ingredient is a statement about $n$ i.i.d. copies of $O = (U, \Delta, W)$. Neither $T \perp C \mid W$ nor predictability of $W$ is used anywhere, which is precisely why Lemma 3.1 does not obstruct the argument.

**Removing the horizon**

The closed forms of Section 4 integrate over the whole half-line, so consistency to $\theta_\tau^\dagger$ is not yet consistency to $\theta^\dagger$. Assume

$\sup_{\theta\in\Theta} \int_\tau^\infty [\mathbb{E}(|W| \mid U = t, \Delta = 1) + |\mu_\theta(t)|]\, F_1(\mathrm{d}t) \to 0$ as $\tau \to \infty$.

**Lemma 3.12.** Under (T), $\Psi_\tau \to \Psi$ uniformly on compact subsets of $\Theta_{\max}$. If in addition $\Psi$ is strictly decreasing with a finite root $\theta^\dagger$, then $\theta_\tau^\dagger \to \theta^\dagger$.

*Proof.* For the second claim, fix $\varepsilon > 0$. Strict monotonicity gives $\Psi(\theta^\dagger - \varepsilon) > 0 > \Psi(\theta^\dagger + \varepsilon)$; uniform convergence transfers both strict inequalities to $\Psi_\tau$ for all large $\tau$, and the root of $\Psi_\tau$ lies between them.

Lemma 3.12 concerns *population* quantities only. It does not deliver consistency of the untruncated estimator, because that additionally requires

$\lim_{\tau\to\infty} \limsup_{n\to\infty} \mathbb{P}\big(\sup_{\theta\in\Theta}|\Psi_n(\theta) - \Psi_{n,\tau}(\theta)| > \varepsilon\big) = 0$ for every $\varepsilon > 0$,

which does not follow from (T). The obstruction is concrete: beyond $\tau$ the risk sets are small, and although each summand $W_i - \mathbb{S}_n^{(1)}/\mathbb{S}_n^{(0)}$ is bounded by the range of $W$ within the risk set, that range grows with $n$, so the product of a vanishing event fraction and a growing range need not vanish uniformly.

**Theorem 3.13 (Fixed-horizon consistency in the memoryless benchmark).** Let $U \sim \mathrm{Exp}(\kappa)$ and let (C1) – (C3) and Corollary 3.9 hold. Then $\theta_\tau^\dagger = \theta^\dagger$ for every $\tau$, since by (4) the horizon enters $\Psi_\tau$ only through the strictly positive scalar $F_1([0,\tau])$. Consequently Theorem 3.11 gives $\hat{\theta}_{n,\tau} \xrightarrow{p} \theta^\dagger$ for every fixed $\tau$: truncation does not move the estimand, so it costs nothing, and neither Lemma 3.12 nor (E-T) is needed.

The title is deliberately *fixed-horizon*, not *removal of the horizon*. Theorem 3.13 concerns $\hat{\theta}_{n,\tau}$ and says nothing yet about $\hat{\theta}_n$, the estimator a Cox routine returns from all event times — and that is the object the influence function, the exact $J$ and $K$, and the software comparisons of Section 6 refer to.

Equality of population roots removes truncation bias from the estimand; it does not control the random tail of the empirical score.

**Reduction of (E-T) by self-similarity**

The reduction below requires more than $U \sim \mathrm{Exp}(\kappa)$: it requires the *marked* law to be memoryless,

$$(A,\, U-\tau,\, \Delta) \mid \{U > \tau\} \overset{d}{=} (A,\, U,\, \Delta). \qquad (5)$$

Exponentiality of $U$ alone does not give (5). If $\lambda_T(t) + \lambda_C(t) \equiv \kappa$ with time-varying components, then $U \sim \mathrm{Exp}(\kappa)$ exactly, but $\mathbb{P}(\Delta = 1 \mid U = t) = \lambda_T(t)/\kappa$ drifts, and so does the mark law in the tail: with $\kappa = 1$ and $\lambda_T(t) = 1/(1+t)$ one has $\mathbb{P}(\Delta = 1) = 0.5963$ while $\mathbb{P}(\Delta = 1 \mid U > 1) = 0.3613$ and $\mathbb{P}(\Delta = 1 \mid U > 4) = 0.1704$. We therefore state the reduction for the *competing-exponential benchmark*, $T \sim \mathrm{Exp}(\lambda)$ and $C \sim \mathrm{Exp}(\mu)$ independent, where $\Delta \sim \mathrm{Bernoulli}(p)$ with $p = \lambda/q$, $q = \lambda + \mu$, and $\Delta \perp U$. Corollary 3.5(ii) and the identity $\theta_\tau^\dagger = \theta^\dagger$ need only $U \sim \mathrm{Exp}(\kappa)$, since $R_\theta$ depends on the law of $U$ alone: it is the reduction, not the estimand, that needs the stronger hypothesis.

Let $M_\tau = \#\{i : U_i > \tau\}$. Every event with $U_i > \tau$ has a risk set contained in $\{j : U_j > \tau\}$, so the discarded part of the score depends on the tail subjects alone, and since $W_i = \gamma\tau + \{A_i + \gamma(U_i - \tau)\}$ the common shift $\gamma\tau$ cancels from every difference $W_i - \mu_\theta(U_i)$ and from every ratio of exponential weights. Hence, exactly,

$$\Psi_n(\theta) - \Psi_{n,\tau}(\theta) = \frac{M_\tau}{n}\, \widetilde{\Psi}_{M_\tau}(\theta), \qquad (6)$$

where $\widetilde{\Psi}_m$ is the *same* observed-data score functional, event indicators included, evaluated on $m$ i.i.d. observations from the *same* law. The tail of the problem is a scaled copy of the problem.

**Proposition 3.14 (Reduction, competing-exponential benchmark).** If

$$B := \sup_{m \ge 1} \mathbb{E}\left[\sup_{\theta \in \Theta} |\widetilde{\Psi}_m(\theta)|\right] < \infty,$$

then (E-T) holds, and consequently $\hat{\theta}_n \overset{p}{\to} \theta^\dagger$.

*Proof.* Conditioning on $M_\tau$ and using that, given $M_\tau = m$, the tail sample is an i.i.d. sample of size $m$ from the original law,

$$\mathbb{E}\left[M_\tau/n \sup_\theta |\widetilde{\Psi}_{M_\tau}|\right] = \mathbb{E}\left[M_\tau/n\ \mathbb{E}\{\sup_\theta |\widetilde{\Psi}_{M_\tau}| \mid M_\tau\}\right] \le B\, \mathbb{E}\, M_\tau/n = B\, e^{-q\tau},$$

since $M_\tau \sim \mathrm{Binomial}(n, e^{-q\tau})$. Markov's inequality gives $\mathbb{P}\big(\sup_\theta |\Psi_n - \Psi_{n,\tau}| > \varepsilon\big) \le B\varepsilon^{-1} e^{-q\tau}$, uniformly in $n$; letting $\tau \to \infty$ yields (E-T).

The reduction is not circular: it asks only that $\widetilde{\Psi}_m$ be bounded in $L^1$, not that it converge.

**Lemma 3.15 (Uniform $L^1$ bound).** Let $\Theta = [\theta_-, \theta_+] \Subset \Theta_{\max}$ and suppose that, for both endpoints $\vartheta \in \{\theta_-, \theta_+\}$,

$$\mathbb{E}W^2 < \infty, \qquad \mathbb{E}\big[|W| e^{\vartheta W}\big] < \infty, \qquad \mathbb{E}\big[|A| e^{\vartheta A}\big] < \infty.$$

Then $B < \infty$.

*Proof. Step 1: the supremum is attained at an endpoint.* For every realised sample the empirical score is non-increasing, $\widetilde{\Psi}_m{}'(\theta) = -m^{-1}\sum_{i:\Delta_i=1} \hat{v}_i\,(\theta) \le 0$, each $\hat{v}_i$ being an empirical tilted risk-set variance. Hence $\sup_{\theta\in\Theta}\left|\widetilde{\Psi}_m(\theta)\right| = \max\{\left|\widetilde{\Psi}_m(\theta_-)\right|, \left|\widetilde{\Psi}_m(\theta_+)\right|\}$, and it suffices to bound $\mathbb{E}\left|\widetilde{\Psi}_m(\theta)\right|$ at a fixed $\theta$.

*Step 2: each risk-set size contributes boundedly.* Consider an exit occurring when the risk set contains $k$ subjects. For $k = 1$ the exiting subject is the whole risk set and $R_1(\theta) = 0$; assume $k \ge 2$. Subtracting the current time, memorylessness gives $X_0 = A_0$ for the exiting subject and $X_j = A_j + \gamma E_j$ with $E_j \overset{iid}{\sim} \operatorname{Exp}(q)$ for the $k-1$ others. With $\bar{X}_{k,\theta}$ the $e^{\theta\cdot}$-weighted mean of $X_0, \dots, X_{k-1}$ and $R_k(\theta) = X_0 - \bar{X}_{k,\theta}$,

$$\mathbb{E}\left|\widetilde{\Psi}_m(\theta)\right| \le \frac{p}{m}\sum_{k=1}^{m} \mathbb{E}\,|R_k(\theta)|,$$

so it suffices that $\sup_k \mathbb{E}|R_k(\theta)| < \infty$. Put $Y_j = e^{\theta X_j}$, choose $c > 0$ with $\alpha = \mathbb{P}(Y_j \ge c) > 0$, and let $N_k = \sum_{j<k} \mathbf{1}\{Y_j \ge c\}$. On $\{N_k \ge \alpha(k-1)/2\}$ the denominator is at least $c\alpha(k-1)/2$, so

$$\left|\bar{X}_{k,\theta}\right| \le \frac{2\left(|X_0|e^{\theta X_0} + \sum_{j<k}|X_j|\,e^{\theta X_j}\right)}{c\alpha(k-1)},$$

whose expectation is $O(1)$ because the numerator has expectation $O(k)$. On the complement, Chernoff's bound gives $\mathbb{P}(N_k < \alpha(k-1)/2) \le e^{-c_0 k}$; since a weighted mean lies between the extreme values, $\left|\bar{X}_{k,\theta}\right| \le \max_{j<k}\left|X_j\right|$ and $\mathbb{E}\max_{j<k}\left|X_j\right|^2 \le \sum_{j<k} \mathbb{E}\,X_j^2 = O(k)$, so by Cauchy–Schwarz the bad event contributes $O\left(\sqrt{k}\,e^{-c_0 k/2}\right)$. Both terms are bounded uniformly in $k$.

The essential point is that $\mu_\theta$ is a tilted *average*, not an extreme: bounding $|W_k - \mu_\theta(U_k)|$ by the range of $W$ within the risk set would give only $O(\log m)$, which is not enough.

**Remark 3.16 (Numerical check of the $L^1$ bound).** At $\gamma = q = \sigma_A = 1$, $\theta^\dagger = -0.618034$, $\Theta = [\theta^\dagger - 0.25, \theta^\dagger + 0.25]$, no censoring:

| $m$ | $10$ | $10^2$ | $10^3$ | $10^4$ | $10^5$ |
|---|---|---|---|---|---|
| $\mathbb{E}\,\widetilde{\Psi}_m(\theta^\dagger)$ | $-0.039$ | $-0.010$ | $-0.005$ | $-0.004$ | $0.000$ |
| $\mathbb{E}\sup_\Theta\left|\widetilde{\Psi}_m\right|$ | 0.430 | 0.413 | 0.381 | 0.363 | 0.363 |
| $\mathbb{E}\left[m^{-1}\sum_i \lvert r_i\rvert\right]$ | 0.596 | 0.768 | 0.792 | 0.796 | 0.798 |

The middle row is the quantity entering $B$, computed at the two endpoints by Step 1; it is flat near 0.363. The third row is the mean absolute score *contribution*, which dominates $\left|\widetilde{\Psi}_m\right|$ and is the sharper diagnostic; it is flat near 0.798 across four orders of magnitude while $\log m$ grows fourfold.

**Remark (Separation).** The localisation in Theorem 3.11 also bounds the probability of separation, since

$$\{\text{separation}\} \subseteq \{\Psi_{n,\tau}(\theta_-) \le 0\} \cup \{\Psi_{n,\tau}(\theta_+) \ge 0\},$$

whose probability tends to zero. The inclusion is one-directional and should not be read as an identity: failure of the score to change sign on $[\theta_-, \theta_+]$ may also mean that a finite root exists outside that interval. Separation is a finite-$n$ phenomenon of the strong-contamination regime, not an obstruction to the asymptotics.

Condition (T) is not restrictive away from the memoryless case: it holds whenever $\mathbb{E}(|W| \mid U = t, \Delta = 1)$ grows at most polynomially while $F_1$ has an exponential tail, which covers Weibull failure times with exponential or administrative censoring. The general non-memoryless case, requiring (T) and (E-T) jointly, is treated in Supplementary Material.

# Continuous-time geometry under affine follow-up contamination

Corollary 3.5 gives the exact least-false coefficient. We now study its scale-free geometry, the way it depends on the law of the baseline covariate, and what survives when the memoryless benchmark is left.

## Gaussian–exponential benchmark: dimensionless reduction

Throughout this subsection $A \sim N(0, \sigma_A^2)$ and $U \sim \mathrm{Exp}(\kappa)$. Define the contamination ratio and the standardised coefficient

$$\rho = \frac{\gamma}{\kappa \sigma_A}, \qquad z = -\sigma_A \theta^\dagger.$$

**Proposition 4.1 (Scale-free reduction).** The map $\rho \mapsto z$ is

$$z(\rho) = \frac{\sqrt{1+4\rho^2} - 1}{2\rho} = \frac{2\rho}{1 + \sqrt{1+4\rho^2}}, \qquad \text{with inverse} \qquad \rho = \frac{z}{1 - z^2}, \tag{7}$$

and it satisfies

1. $0 < z(\rho) < 1$ and $z'(\rho) > 0$ for all $\rho > 0$;
2. $z(\rho) = \rho - \rho^3 + 2\rho^5 + O(\rho^7)$ as $\rho \downarrow 0$;
3. $z(\rho) = 1 - \frac{1}{2\rho} + \frac{1}{8\rho^2} + O(\rho^{-4})$ as $\rho \to \infty$.

Equivalently, $\theta^\dagger \sim -\gamma/(\kappa \sigma_A^2)$ as $\rho \downarrow 0$, while $\theta^\dagger \to -1/\sigma_A$ as $\rho \to \infty$.

The second form in (7) is the numerically stable one: the first loses all significance through cancellation as $\rho \downarrow 0$.

The content of Proposition 4.1 is a saturation statement. Contamination enters only through the single ratio $\rho$, so the two natural scales — how fast subjects leave, $1/\kappa$, and how variable the honest covariate is, $\sigma_A$ — act only in the combination $\gamma/(\kappa\sigma_A)$. And the raw least-false coefficient does not grow without bound as contamination strengthens: it saturates at $-1/\sigma_A$, a scale fixed by the baseline covariate alone. Doubling $\gamma$ in an already heavily contaminated design changes almost nothing; doubling it in a weakly contaminated one nearly doubles the bias.

Note also what *is* unbounded. The saturation is of $\theta^\dagger$, the coefficient as reported. It says nothing about the association the covariate expresses once its own spread is accounted for, since $\mathrm{SD}(W)$ grows linearly in $\gamma$; that comparison belongs with the grouped-time results and is taken up in Section 5.

## Beyond Gaussian covariates

The Gaussian closed form is not distribution-free. For a general honest covariate, the strong-contamination limit depends on its standardized distribution through the root derived in the Supplementary Material, rather than on the variance alone. The Gaussian value is therefore a benchmark, not a universal constant.

## Administrative censoring: failure of the scale reduction

Let $T \sim \mathrm{Exp}(\lambda)$ and $C \equiv c$, so $U = T \wedge c$, and define the general contamination ratio $\rho_c = \gamma \mathrm{SD}(U)/\sigma_A$. The residual-life representation (2) remains available in closed form even though its root does not. Writing $a_\theta = \lambda - \theta\gamma$ and $d = c - t$, the tilted residual $U - t$ given $U \geq t$ is $\min(E, d)$ with $E \sim \mathrm{Exp}(\lambda)$ tilted by $e^{\theta\gamma \cdot}$, whence for $0 \leq t < c$

$$R_\theta(t) = \frac{\frac{\lambda}{a_\theta}\{1 - e^{-a_\theta d}\} - \theta\gamma\, d\, e^{-a_\theta d}}{\lambda - \theta\gamma\, e^{-a_\theta d}}, \qquad \Psi_c(\theta) = -K_A{}'(\theta)\left(1 - e^{-\lambda c}\right) - \gamma \int_0^c R_\theta(t)\, \lambda e^{-\lambda t} \mathrm{d}t. \tag{10}$$

At $\theta = 0$ this reduces to $R_0(t) = \left(1 - e^{-\lambda d}\right)/\lambda = \mathbb{E}\min(E, d)$, as it must. The root of $\Psi_c$ is obtained by one-dimensional quadrature.

Two facts follow, and only these two should be read off.

*$\rho$ is no longer a sufficient summary.*

In the Gaussian memoryless benchmark, constancy of $R_\theta$ in $t$, together with the quadratic cumulant generating function $K_A$, is exactly what reduces the root to the single dimensionless ratio $\rho$; for general $A$ a standardised form of the whole law of $A$ is needed as well, as Section 4.2 showed. Under administrative censoring $R_\theta(t)$ depends on $t$ through $d = c - t$: subjects near the administrative boundary have almost no residual life left to accumulate. The root of $\Psi_c$ therefore depends on the entire family $\{\mathcal{L}(U - t \mid U \geq t)\}_{t \leq c}$, not on any single moment of it, and two censoring laws matched on $\rho$ need not produce the same $\theta^\dagger$. Concretely, at $\rho = 1$ the memoryless benchmark gives $\theta^\dagger = -0.618034$, whereas $T \sim \mathrm{Exp}(1)$ with $C \equiv 1$ and $\gamma = 1/\mathrm{SD}(U) = 2.785247$, which has the same $\rho_c = 1$, gives $\theta^\dagger = -0.823151$: a 33% difference at matched contamination ratio. A stronger version of this reappears in Section 6, where changing the censoring law can reverse even the sign of $K - J$ at matched $(p, \rho)$.

**The same saturation level is reached by a boundary-layer argument.**

Administrative censoring is not memoryless, so the limit equation (9) of Section 4.2 does not transfer automatically. It nevertheless holds.

**Lemma 4.3 (Large-contamination limit under administrative censoring).** Fix $\theta < 0$. Then $\gamma \int_0^c R_\theta(t) \lambda e^{-\lambda t} \mathrm{d}t \to -\left(1 - e^{-\lambda c}\right)/\theta$ as $\gamma \to \infty$, so the limiting score equation is again $\theta K_A{}'(\theta) = 1$.

*Proof.* Write $\theta = -r$ with $r > 0$, $a = \lambda + r\gamma$ and $d = c - t$. The transition region is *not* of width $O(\gamma^{-1})$: at $d_\gamma = \log\gamma/(2r\gamma)$ one has $e^{-ad_\gamma} \asymp \gamma^{-1/2}$ and, substituting in (10), $\gamma R_\theta(c - d_\gamma) \asymp \log\gamma/(2r)$ rather than $1/r$. The split must therefore be taken at a logarithmically inflated scale.

Fix $M > 1/r$ and set $L_\gamma = M\log\gamma$. On $d \in [0, L_\gamma/\gamma]$ use the crude bound $R_\theta(t) \le \mathbb{E}\min(E, d) \le d$, so that

$$\gamma \int_{c-L_\gamma/\gamma}^{c} R_\theta(t)\, \lambda e^{-\lambda t} \mathrm{d}t \ \le\ \lambda\gamma \int_0^{L_\gamma/\gamma} d\ \mathrm{d}d = \frac{\lambda L_\gamma^2}{2\gamma} = O\left(\frac{(\log\gamma)^2}{\gamma}\right) \to 0.$$

On $d \ge L_\gamma/\gamma$ the remainder is controlled explicitly. Direct algebra in (10) gives the exact identity

$$R_\theta(c-d) - \frac{1}{a} = \frac{e^{-ad}(r\gamma d - 1)}{\lambda + r\gamma e^{-ad}}, \qquad (11)$$

so that $|R_\theta(c-d) - a^{-1}| \le \lambda^{-1}(r\gamma d + 1)e^{-ad}$. Since $L_\gamma/\gamma > 1/a$ for all large $\gamma$, the map $d \mapsto (r\gamma d + 1)e^{-ad}$ is decreasing on $[L_\gamma/\gamma, \infty)$, and $e^{-aL_\gamma/\gamma} \le e^{-rM\log\gamma} = \gamma^{-rM}$, whence

$$\sup_{d \ge L_\gamma/\gamma} \gamma |R_\theta(c-d) - \frac{1}{a}| \ \le\ \frac{\gamma(rM\log\gamma + 1)}{\lambda} \gamma^{-rM} \ \le\ C(1 + \log\gamma)\gamma^{1-rM} \to 0$$

for a constant $C$ not depending on $\gamma$, because $rM > 1$. Hence $\gamma R_\theta(t) = \gamma/a + o(1) \to 1/r$ uniformly on that range. Hence $\gamma \int_0^c R_\theta(t)\lambda e^{-\lambda t}\mathrm{d}t \to r^{-1}(1 - e^{-\lambda c}) = -(1 - e^{-\lambda c})/\theta$. Substituting into (10) leaves $-(1 - e^{-\lambda c})\{K_A'(\theta) - 1/\theta\}$, whose root is $\theta K_A'(\theta) = 1$.

**The trajectory can overshoot that limit.**

For the numerical illustration we return to $A \sim N(0, \sigma_A^2)$, for which Lemma 4.3 gives the limit $-1/\sigma_A$. With $\lambda = 1$, $c = 1$ and $\sigma_A = 1$ the coefficient slightly overshoots that limit before returning to it:

| $\gamma$ | 1 | 10 | 27.83 | 50 | $10^2$ | $10^3$ | $10^5$ |
|---|---|---|---|---|---|---|---|
| $\theta^\dagger$ | −0.3975 | −1.0116 | −1.0238 | −1.0220 | −1.0174 | −1.0049 | −1.0000 |

The maximum of $|\theta^\dagger|$ is 1.023841 at $\gamma = 27.83$, an overshoot of 2.38%, occurring at $\rho_c = 9.99$; thereafter $|\theta^\dagger|$ decreases slowly back to 1. We state this as a numerical observation about this particular benchmark and claim no general theorem: nothing here establishes non-monotonicity for administrative censoring at other $(\lambda, c, \sigma_A)$, still less for other censoring laws.

# Grouped-time geometry and non-commuting limits

Reported exit times are rarely continuous. Renewal dates, reporting cycles and settlement calendars all place $U$ on a grid, and the analyst then fits the Cox model to tied data. This section works out the exact consequence.

## Breslow benchmark and the exact phase equation

Throughout we work in the competing-exponential benchmark of Section 3: $A \sim N(0, \sigma_A^2)$, $T \sim \mathrm{Exp}(\lambda)$ and $C \sim \mathrm{Exp}(\mu)$ independent, with $\kappa = \lambda + \mu$, $U = T \wedge C$ and $\Delta = \mathbf{1}\{T \le C\}$, so that

$$U \sim \mathrm{Exp}(\kappa), \qquad \varDelta \perp U, \qquad \mathbb{P}(\varDelta = 1) = p = \lambda/\kappa,$$

and

$$U_h = h\left\lceil \frac{U}{h} \right\rceil, \qquad W_h = A + \gamma U_h.$$

Exponentiality of $U$ alone would not suffice: as Section 3.6 showed, the mark law $\mathbb{P}(\varDelta = 1 \mid U = t)$ can then drift with $t$, and the event fraction would not factor out. The latent exit time is rounded *up* to the end of its interval, so $U_h/h$ is geometric on $\{1,2,\dots\}$ with success probability $1 - e^{-a}$, $a = \kappa h$. Subjects sharing a value of $U_h$ are tied, and all events at a tied time are given the same risk-set denominator: this is the rule of Breslow (1974), and it is part of the definition of the benchmark, not an approximation to it. The rule of Efron (1977) defines a different population equation and is not treated here. We stress that this is deliberately the naive analysis: when exit times are genuinely grouped, the appropriate model is the discrete-time likelihood of Prentice and Gloeckler (1978), and our object of study is precisely what the continuous-time partial likelihood converges to when it is applied to such data instead. Because $\varDelta \perp U$, the event fraction $p$ factors out of the population score and does not enter any root below. Write

$$x = -\theta_h^{\dagger} > 0, \qquad z = \sigma_A x, \qquad a = \kappa h, \qquad d = \frac{\gamma h}{\sigma_A}.$$

**Proposition 5.1 (Exact Breslow equation).** The grouped least-false coefficient solves

$$\sigma_A^2\, x = \frac{\gamma h}{\exp\{(\kappa + \gamma x)h\} - 1}, \qquad \text{equivalently} \qquad z = \frac{d}{e^{a+dz} - 1}, \tag{12}$$

and for every $a > 0$, $d > 0$ this equation has exactly one positive root.

*Proof.* By memorylessness, given $U_h \geq kh$ the grouped residual satisfies

$$\mathbb{P}(U_h - kh = mh \mid U_h \geq kh) = (1 - e^{-a})e^{-am}, \qquad m = 0,1,2,\dots$$

Tilting by $e^{\theta\gamma U_h}$ with $\theta = -x$ multiplies the $m$th term by $e^{-\gamma xhm}$, so the tilted law is again geometric, with ratio $e^{-(a+\gamma xh)}$ and mean $\{e^{a+\gamma xh} - 1\}^{-1}$. The tilted mean residual life is therefore constant in $k$,

$$R_{-x,h} = \frac{h}{e^{a+\gamma xh} - 1},$$

and Corollary 3.5 with $-K_A{}'(-x) = \sigma_A^2 x$ reduces to $F_1\{\sigma_A^2 x - \gamma R_{-x,h}\} = 0$, which is the first display. Dividing by $\sigma_A$ and using $a = \kappa h$, $d = \gamma h/\sigma_A$ gives the second. For uniqueness put $F(z) = z(e^{a+dz} - 1) - d$. Then $F(0) = -d < 0$, $F(z) \to \infty$, and

$$F'(z) = e^{a+dz} - 1 + dz\, e^{a+dz} > 0$$

for $a > 0$, $z > 0$, so $F$ has exactly one positive zero.

The structural point is a change of dimension. In continuous time the whole problem collapsed to the single ratio $\rho = \gamma/(\kappa\sigma_A)$; here it does not.

> Unlike the continuous Gaussian benchmark, grouped time is governed by the two-dimensional phase pair $(a, d)$, and not by the single ratio $d/a$.

The two coordinates have distinct meanings. Coarseness $a = \kappa h$ measures the grid against the exit-time scale: how many subjects leave within one reporting interval. Contamination $d = \gamma h/\sigma_A$ measures how much terminal-time signal accrues over one interval, relative to the spread of the honest covariate. Two designs with the same $\rho$ but different resolutions sit at different points of the surface and give different answers.

**Remark (Separation on a grid).** Equation (12) is a population statement. At finite $n$, separation is no longer mechanically forced, because ties disrupt the strict terminal ordering that drives it in continuous time; but it remains possible, and whether it occurs depends on the tied-event configuration and on the tie rule.

## Fixed-grid geometry

Fix $a > 0$ and vary $d$. From (12), $z \to 0$ both as $d \downarrow 0$ and as $d \to \infty$, so $|\theta_h^\dagger|$ attains an interior maximum. The substitution $y = dz$ makes this explicit: eliminating $d$ from (12) gives

$$z^2 = g(y), \qquad g(y) = \frac{y}{e^{a+y} - 1}. \tag{13}$$

**Proposition 5.3 (Unique hump; Lambert-$W$ location and height).** For each $a > 0$ the function $g$ has a unique stationary point $y_* \in (0,1)$, a maximum, characterised by $(1-y)e^y = e^{-a}$ and given in closed form by

$$y_* = 1 + W_0\left(-e^{-(1+a)}\right).$$

Consequently

$$\max_{\gamma>0}|\theta_h^\dagger| = \frac{\sqrt{1-y_*}}{\sigma_A}, \qquad \text{attained at} \qquad \gamma_* = \frac{\sigma_A}{h} \cdot \frac{y_*}{\sqrt{1-y_*}}.$$

*Proof.* $g'(y)$ has the sign of $N(y) = e^{a+y}(1-y) - 1$. Since $N'(y) = -y\,e^{a+y} < 0$ for $y > 0$, $N$ is strictly decreasing; $N(0) = e^a - 1 > 0$ and $N(y) \to -\infty$, so $N$ has exactly one positive zero $y_*$, and $g$ increases on $(0, y_*)$ and decreases after: the stationary point is the unique maximum. Writing $u = y - 1$ turns $(1-y)e^y = e^{-a}$ into $ue^u = -e^{-(1+a)}$, and $y_* \in (0,1)$ selects the principal branch, giving the stated $y_*$. At $y_*$ we have $e^{a+y_*} = (1-y_*)^{-1}$, hence $e^{a+y_*} - 1 = y_*/(1-y_*)$ and $g(y_*) = 1 - y_*$, so $z_{\max} = \sqrt{1-y_*}$. Finally $d_* = y_*/z_* = y_*/\sqrt{1-y_*}$, and $\gamma_* = \sigma_A d_*/h$.

| $a = \kappa h$ | 0.05 | 0.2 | 1.0 | 3.0 |
|---|---|---|---|---|
| $y_*$ | 0.28381054 | 0.50676058 | 0.84140566 | 0.98133937 |
| $\sigma_A \max\lvert\theta_h^\dagger\rvert = \sqrt{1-y_*}$ | 0.84627977 | 0.70231006 | 0.39823905 | 0.13660391 |
| $d_*$ | 0.33536 | 0.72156 | 2.11282 | 7.18383 |

Direct maximisation of the root of (12) reproduces the second row to eight decimals.

Mechanically: at fixed resolution, increasing contamination first strengthens the spurious coefficient, but eventually makes the grouped terminal component so dominant within tied risk sets that the fitted coefficient shrinks back toward zero. This non-monotonicity should not be identified with the administrative overshoot of Section 4.3. Both are humps; the mechanisms differ. There the cause is a non-constant residual-life profile near a fixed boundary, here it is the competition between grid resolution and contamination strength within tied risk sets.

**Vanishing coefficient, diverging association**

**Proposition 5.4 (Strong contamination at fixed resolution).** Fix $a > 0$. As $d \to \infty$,

$$z(a,d) \sim \frac{W_0(e^{-a}d^2)}{d}, \qquad \text{so} \qquad \theta_h^\dagger \to 0.$$

*Proof.* From (13), $d^2 = y(e^{a+y} - 1)$, which for large $d$ gives $d^2 e^{-a} \sim y e^y$, that is $y \sim W_0(e^{-a}d^2)$; and $z = y/d$.

At $a = 0.2$ the exact roots at $d = 10^2, 10^4, 10^6$ are $7.057 \times 10^{-2}$, $1.548 \times 10^{-3}$, $2.424 \times 10^{-5}$, against the asymptotic $7.056 \times 10^{-2}$, $1.548 \times 10^{-3}$, $2.424 \times 10^{-5}$.

Vanishing of the raw coefficient is not the disappearance of the spurious association. Define, for the first time in this paper, the standardised pseudo-effect

$$\zeta^\dagger = |\theta^\dagger| \mathrm{SD}(W).$$

Since $U_h/h$ is geometric, $\mathrm{SD}(U_h) = h\, e^{-a/2}/(1 - e^{-a})$, so

$$\zeta_h^\dagger = z(a,d)\sqrt{1 + \frac{d^2 e^{-a}}{(1-e^{-a})^2}} \sim \frac{e^{-a/2}}{1 - e^{-a}} W_0(e^{-a}d^2) = O(\log d). \tag{14}$$

In continuous time, writing $z_{\text{cont}}$ for the map $z(\rho)$ of Proposition 4.1 evaluated at $r = \gamma/(\kappa\sigma_A)$,

$$\zeta_{\text{cont}}^\dagger = z_{\text{cont}}(r)\sqrt{1 + r^2} \sim r.$$

> Grouping reverses the asymptotic behaviour of the raw coefficient but does not regularise the standardised pseudo-effect: it lowers the divergence from linear to logarithmic.

A word on what $\zeta^\dagger$ is. It is not a true association and still less a causal effect: in a misspecified model no such object is being estimated. It is the magnitude of the working-model log-hazard contrast associated with a one-standard-deviation difference in $W$, and $\exp(\zeta^\dagger)$ is the corresponding pseudo-hazard ratio — the number a reader of the output would quote.

This resolves what would otherwise look like a paradox. Reading Proposition 5.4 alone, one might conclude that coarse reporting protects against terminal-time contamination. It does not. As $\gamma \to \infty$ the covariate becomes a monotone function of the outcome, and what shrinks is only the coefficient attached to a covariate whose own spread is growing without bound. The comparison $\zeta_{\text{cont}}^\dagger \sim r$ against $\zeta_h^\dagger = O(\log d)$ is the honest one, and it is the content of Panel C below.

## Joint limits and phase diagram

**Corollary 5.5 (Non-commuting limits).**

$$\lim_{\gamma\to\infty} \lim_{h\downarrow 0} \theta_h^\dagger = -\frac{1}{\sigma_A}, \qquad \text{whereas} \qquad \lim_{h\downarrow 0} \lim_{\gamma\to\infty} \theta_h^\dagger = 0.$$

*Proof.* Taking $h \downarrow 0$ first recovers the continuous benchmark, whose limit is $-1/\sigma_A$ by Proposition 4.1; taking $\gamma \to \infty$ first gives 0 at every fixed $h$ by Proposition 5.4, and 0 is unchanged by $h \downarrow 0$.

This is the conceptual centre of the section. There is no such thing as “the” strong-contamination behaviour of the grouped estimator: the answer depends on the order in which resolution and contamination are taken to their limits, so any asymptotic statement must say which regime it describes.

Two intermediate regimes interpolate, and they must not be confused with one another.

**Proposition 5.6 (Joint edge).** Let $h \downarrow 0$ and $\gamma \to \infty$ jointly with $\gamma h/\sigma_A \to d \in (0, \infty)$. Then $a \to 0$, the contamination coordinate converges to $d$, and $z$ converges to $z_J(d)$, the unique positive root of

$$z_J = \frac{d}{e^{dz_J} - 1}. \qquad (15)$$

*Proof.* Let $a_n \to 0$, $d_n \to d > 0$, and let $z_n$ solve $F_n(z) = z(e^{a_n + d_n z} - 1) - d_n = 0$. Then $F_n(0) = -d_n < 0$ and $F_n(1) = e^{a_n + d_n} - 1 - d_n > 0$, since $e^u > 1 + u$ for $u > 0$; hence $0 < z_n < 1$ for all $n$. On $[0,1]$ the functions $F_n$ converge uniformly to $F(z) = z(e^{dz} - 1) - d$, which is strictly increasing with the unique positive root $z_J(d)$. Every convergent subsequence of the bounded sequence $(z_n)$ therefore has limit $z_J(d)$, and so $z_n \to z_J(d)$.

**Proposition 5.7 (Directional limits at the singular origin).** Let $a_n \downarrow 0$ and $d_n \downarrow 0$ with $d_n/a_n \to r \in [0, \infty]$. Then

$$z(a_n, d_n) \to z_0(r) = \frac{\sqrt{1 + 4r^2} - 1}{2r}, \qquad (16)$$

with the continuous extensions $z_0(0) = 0$ and $z_0(\infty) = 1$.

*Proof.* As in Proposition 5.6, $0 < z_n < 1$. Expanding $e^{a_n + d_n z} - 1 = (a_n + d_n z)\{1 + O(a_n + d_n)\}$ in (12) gives

$$z_n(a_n + d_n z_n) = d_n\{1 + o(1)\}. \qquad (17)$$

Three cases. If $0 < r < \infty$, divide (17) by $a_n$ and pass to the limit along any convergent subsequence: $z + rz^2 = r$, whose unique root in $(0,1)$ is $z_0(r)$ as displayed. If $r = 0$, dividing by $a_n$ gives $z_n + (d_n/a_n)z_n^2 = (d_n/a_n)\{1 + o(1)\}$, and since $d_n/a_n \to 0$ and $z_n$ is bounded, $z_n \to 0$. If $r = \infty$, divide (17) by $d_n$ instead: $z_n^2 + (a_n/d_n)z_n = 1 + o(1)$ with $a_n/d_n \to 0$ and $0 < z_n < 1$, whence $z_n \to 1$.

The origin $(a, d) = (0,0)$ is therefore *singular*: the limit exists along every ray but depends on which ray is taken, and a single geometric point carries a continuum of limits. Note also that $z_0$ is exactly the continuous-time map $z(\rho)$ of Proposition 4.1, with $r$ in place of $\rho$: approaching the origin along a fixed direction reproduces continuous time, as it should.

These two limits are distinct and should not be conflated. Proposition 5.6 sends $a \downarrow 0$ at *fixed* $d > 0$ and yields a curve in $d$; Proposition 5.7 sends $a$ and $d$ jointly to zero at fixed ratio $d/a$ and yields a family indexed by that ratio. Numerically, at $a = 10^{-6}$ the edge equation (15) reproduces the exact root to seven decimals ($d = 2$: 0.68371052 against 0.68371019), while along $d/a = r$ the exact roots approach $z_0(r)$ from below ($r = 1$: 0.614428, 0.617998, 0.618034 at $a = 10^{-2}, 10^{-4}, 10^{-6}$, against $z_0(1) = 0.61803399$).

### The phase diagram

Figure 1 collects the geometry. Panel A shows the exact two-parameter surface through fixed-$a$ sections together with the joint edge $a \downarrow 0$ at fixed $d$; Panel B shows the directional limits at the singular origin; Panel C compares the standardised pseudo-effect in continuous and grouped time.

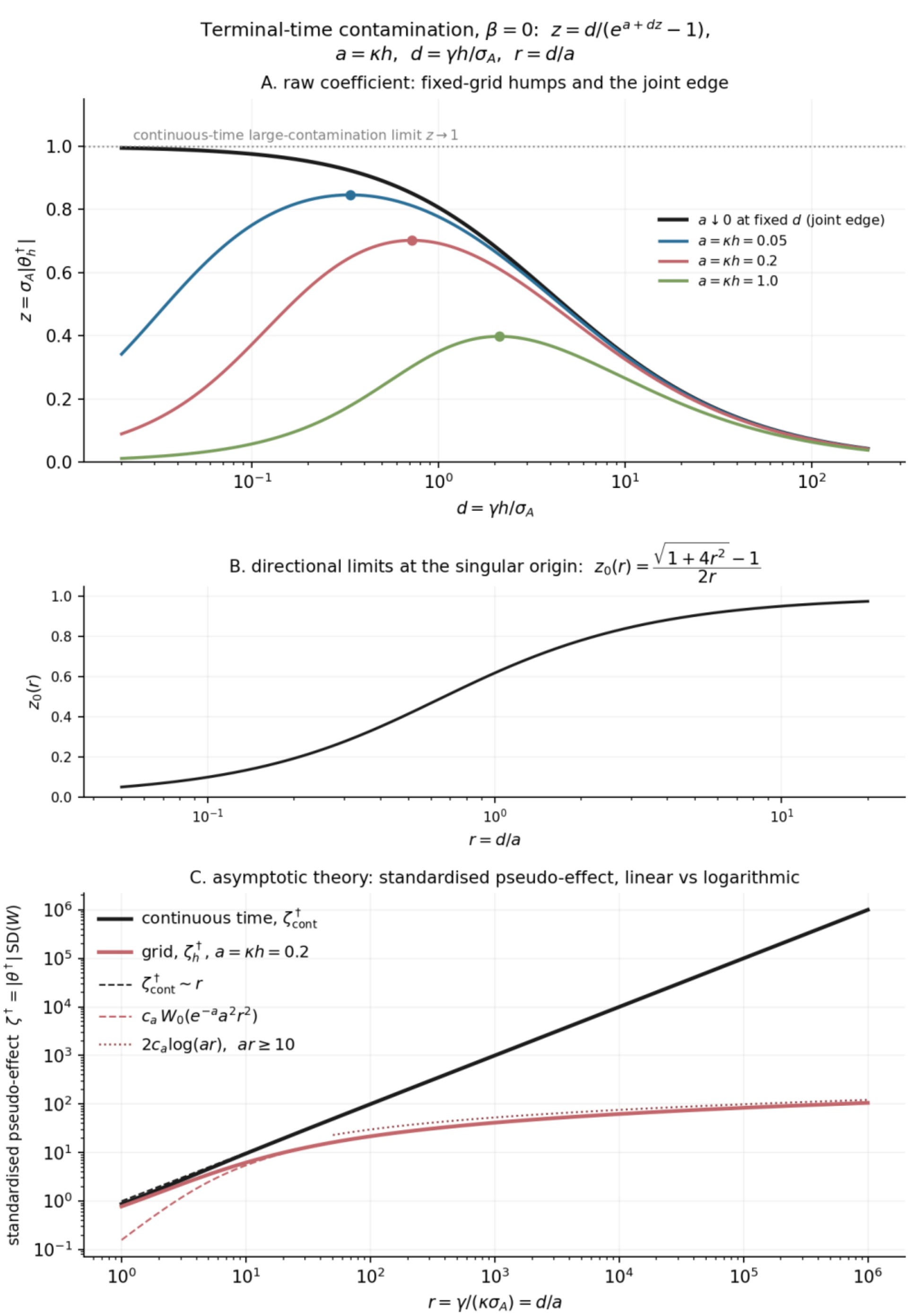

*Terminal-time contamination at $\beta = 0$, with $a = \kappa h$, $d = \gamma h/\sigma_A$ and $r = d/a = \gamma/(\kappa\sigma_A)$.* ***Panel A*** *plots $z = \sigma_A|\theta_h^\dagger|$, the root of $z = d/(e^{a+dz} - 1)$: coloured curves are sections at fixed $a$, each with its Lambert-W hump (Proposition 5.3); the black curve is the joint edge $a \downarrow 0$ at fixed $d$ (Proposition 5.6).* ***Panel B*** *shows the directional limits $z_0(r)$ at the singular origin (Proposition 5.7); note that this is a different limit from the joint edge of Panel A.* ***Panel C*** *plots the standardised pseudo-effect $\zeta^\dagger =$*

*$|\theta^\dagger| SD(W)$, linear in $r$ in continuous time against logarithmic on a fixed grid; the Lambert-$W$ and logarithmic grid approximations use $c_a = e^{-a/2}/(1-e^{-a})$. Gridded curves assume ceiling aggregation $U_h = h\lceil U/h \rceil$, independent exponential event and censoring times with total exit rate $\kappa$, Gaussian $A$, additive $W = A + \gamma U_h$, and Breslow handling of tied events. In the continuous exponential benchmark $r$ coincides with the contamination ratio $\rho$; on the grid it does not, the actual ratio being $\rho_h = r\, a e^{-a/2}/(1-e^{-a})$.*

With the geometry of the least-false coefficient complete, the question changes. Sections 3–5 answer *which* spurious target is estimated; Section 6 asks what sampling uncertainty surrounds that target, and finds that the answer reported by default software is wrong in a direction that itself depends on the censoring regime.

# Inference for the least-false target

Sections 3–5 establish *which* target the naive analysis estimates. This section asks how uncertain that estimate is, and finds that the answer printed by default need not be correct and can fail in either direction, with accidental equality on a non-trivial phase boundary.

## Observed-data influence function

We derive the asymptotic distribution as an observed-data $M$-estimation result. This is not a stylistic choice. The standard route to Cox asymptotics, including the robust variance of Lin and Wei (1989), proceeds through a martingale representation that requires $W$ to be predictable and censoring to be conditionally independent given the fitted covariate; Lemma 3.1 shows the second fails and the discussion after it shows the first fails too. The conclusion nevertheless survives, because a $Z$-estimator of an i.i.d. empirical criterion needs neither. The relationship to Cox score residuals is recorded afterwards, once the derivation stands on its own.

**Theorem 6.1 (Asymptotic normality on a finite horizon).** Fix $\tau$ and assume (C1)–(C4), the last as stated in Supplementary Material, together with $J_\tau > 0$. Then

$$\sqrt{n}\left(\hat{\theta}_{n,\tau} - \theta_\tau^\dagger\right) = J_\tau^{-1} \frac{1}{\sqrt{n}} \sum_{i=1}^{n} \varphi_{\tau,\theta_\tau^\dagger}(\mathcal{O}_i) + o_p(1) \Rightarrow N\left(0, \frac{K_\tau}{J_\tau^2}\right),$$

where

$$\varphi_{\tau,\theta}(\mathcal{O}) = \Delta \mathbf{1}\{U \le \tau\}\{W - \mu_\theta(U)\} - e^{\theta W} \int_{[0, U\wedge\tau]} \frac{W - \mu_\theta(t)}{s^{(0)}(\theta, t)} F_1(\mathrm{d}t), \tag{18}$$

and

$$J_\tau = -\Psi_\tau{}'\left(\theta_\tau^\dagger\right) = \int_{[0,\tau]} v_{\theta_\tau^\dagger}(t)\, F_1(\mathrm{d}t), \qquad K_\tau = \mathbb{E}\left[\varphi_{\tau,\theta_\tau^\dagger}(\mathcal{O})^2\right].$$

The proof is an ordinary $Z$-estimator expansion of a smooth functional of the empirical measure and is given in Supplementary Material.

**Corollary 6.2 (Full horizon in the competing-exponential benchmark; conditional).** In the competing-exponential benchmark, *assuming* the $L^2$ tail conditions (A.1)–(A.3) of Supplementary Material, which we state but do not prove, $\varphi_{\tau,\theta_\tau^\dagger} \to \varphi_{\theta^\dagger}$ in $L^2$, $J_\tau \to J$, $K_\tau \to K$, and

$$\sqrt{n}(\hat{\theta}_n - \theta^\dagger) \Rightarrow N\left(0, \frac{K}{J^2}\right),$$

with $\varphi_{\theta^\dagger}$, $J$ and $K$ obtained from (18) by setting $\tau = \infty$.

The separation matters for the same reason it did in Section 3: conditions (C1)–(C3) were stated at a fixed horizon, and (C2) bounds the risk-set denominator only on $[0, \tau]$. Section 3 removed the horizon for consistency by an $L^1$ tail bound; a central limit theorem needs the corresponding $L^2$ control, which does not follow from it.

The two terms of (18) have distinct origins and the second is the one that is easy to miss. The first is the subject's own event contribution. The second says that a subject also enters, through the denominator $\mathbb{S}_n^{(0)}$, every risk set at which it is still alive: perturbing one observation moves not only its own summand but every earlier one. Because $\Psi_n$ is a ratio functional of the empirical measure rather than a plain average, that second channel contributes to the influence function, and it is essential to the sandwich meat. The model-based variance of Section 6.2 does *not* arise by deleting this term — a correctly specified Cox model has the same risk-set contribution in its score residuals — but by replacing the actual variance $K$ of the influence function with the curvature $J$, through the information equality.

**Remark 6.3 (Relation to Cox score residuals).** Written empirically, (18) becomes

$$\hat{\varphi}_i = \Delta_i\{W_i - \hat{\mu}(U_i)\} - e^{\hat{\theta} W_i} \sum_{\substack{j:\Delta_j=1 \\ U_j \le U_i}} \frac{W_i - \hat{\mu}(U_j)}{\sum_k \mathbf{1}\{U_k \ge U_j\} e^{\hat{\theta} W_k}},$$

where

$$\hat{J} = \frac{1}{n} \sum_{j:\Delta_j=1} \hat{v}_{\hat{\theta}}(U_j),$$

coincides with the subject-level Cox score residual. Two scalings must be kept apart: $\hat{J}^{-1}\hat{\varphi}_i$ is the estimated influence value, whereas the case perturbation on the DFBETA scale is $n^{-1}\hat{J}^{-1}\hat{\varphi}_i$, since the total observed information is $n\hat{J}$; implementations may also reverse the sign, according to whether DFBETA is defined as full minus deleted or the reverse. The agreement is algebraic and worth stating, but it is a consequence of the derivation above, not a licence to import the martingale theory that ordinarily justifies those residuals.

## Model-based versus sandwich variance

Throughout this subsection the horizon subscript is suppressed: unless explicitly stated otherwise, $\hat{\theta}$, $J$, $K$ and $\varphi$ denote their fixed-horizon counterparts $\hat{\theta}_{n,\tau}$, $J_\tau$, $K_\tau$ and $\varphi_{\tau,\theta_\tau^\dagger}$. Their full-horizon interpretation is conditional on Corollary 6.2.

Standard Cox output reports the inverse observed information,

$$\widehat{\mathrm{Var}}_{\mathrm{model}}(\hat{\theta}) = \frac{1}{n\hat{J}},$$

which is the correct asymptotic variance only if the information equality $K = J$ holds. That equality is a consequence of correct specification and has no reason to survive here. The $M$-estimation variance is instead

$$\mathrm{Var}_{\mathrm{true}}(\hat{\theta}) \sim \frac{K}{nJ^2}, \qquad \widehat{\mathrm{Var}}_{\mathrm{sand}} = \frac{\hat{K}}{n\hat{J}^2}, \qquad \hat{K} = \frac{1}{n}\sum_{i=1}^{n} \hat{\varphi}_i^2.$$

By Theorem 6.1 and the consistency of $\hat{J}$ and $\hat{K}$, the sandwich estimator is consistent for the true sampling variance under i.i.d. sampling, a finite interior root and a finite second moment of the influence function — and, again, without $T \perp C \mid W$ or predictability of $W$.

## The exact uncensored Gaussian benchmark

The population formulas below are exact. Their interpretation as the asymptotic variance and coverage of the *untruncated* Cox estimator is conditional on Corollary 6.2; the corresponding fixed-horizon influence-function and sandwich theory is unconditional.

Take $A \sim N(0, \sigma_A^2)$, $U \sim \mathrm{Exp}(\kappa)$ and no censoring, and write $z = -\sigma_A \theta^\dagger = z(\rho)$ as in Proposition 4.1. At the root the own-event residual collapses: $\mu_{\theta^\dagger}(t) = \sigma_A^2 \theta^\dagger + \gamma t + \gamma/(\kappa - \gamma\theta^\dagger)$, and the root equation kills the two terms that do not depend on $t$, leaving

$$W - \mu_{\theta^\dagger}(U) = A. \qquad (19)$$

This is what makes the benchmark solvable in closed form.

**Theorem 6.4 (Exact bread and meat).** In the uncensored Gaussian benchmark,

$$J = \sigma_A^2(1 + z^2), \qquad K = \sigma_A^2\, \mathcal{K}(z), \qquad \mathcal{K}(z) = \frac{e^{z^2}(4z^6 + 4z^4 + 6z^2 + 2)}{(1 + z^2)^3} - 1.$$

The bread is immediate. Since $A \perp U$ and the tilt factorises, the tilted risk-set variance splits, and the tilted residual life is exponential with rate $\kappa - \gamma\theta$, so

$$v_\theta(t) = \sigma_A^2 + \frac{\gamma^2}{(\kappa - \gamma\theta)^2} \qquad \text{for every } t.$$

At the root, $\gamma/(\kappa - \gamma\theta^\dagger) = -\sigma_A^2\theta^\dagger$, whence $v_{\theta^\dagger} \equiv \sigma_A^2 + \sigma_A^4(\theta^\dagger)^2 = \sigma_A^2(1 + z^2)$, and $J$ follows because $F_1$ is a probability measure here. The meat requires the full second moment of (18) and is computed in Supplementary Material.

**Theorem 6.5 (Failure of the information equality).** $\mathcal{K}(z) > 1 + z^2$ for every $z \in (0,1)$; equivalently $K > J$ for every $\gamma > 0$.

*Proof.* Put $x = z^2 \in (0,1)$. The claim is $D(x) > 0$ for $x > 0$, where

$$D(x) = e^x(4x^3 + 4x^2 + 6x + 2) - (1 + x)^3(2 + x).$$

The coefficient of $x^m$ in the first term is $c_m = 2/m! + 6/(m-1)! + 4/(m-2)! + 4/(m-3)!$, with terms of negative factorial index read as zero, while $(1+x)^3(2+x) = 2 + 7x + 9x^2 + 5x^3 + x^4$ has degree four. Hence

$$D(x) = x + 2x^2 + 19/3\, x^3 + 73/12\, x^4 + \sum_{m \geq 5} c_m\, x^m,$$

where the first four coefficients are obtained by direct subtraction and every $c_m$ with $m \geq 5$ is positive because nothing is subtracted beyond degree four. All coefficients are strictly positive, so $D(x) > 0$ for $x > 0$.

The last step is worth stating explicitly rather than by an appeal to "the remaining coefficients are also positive": the argument works precisely because the subtracted polynomial has finite degree.

**Remark 6.6 (Numerical confirmation).** Computing $J$ and $K$ from (18) on samples of size $1.5 \times 10^6$ at the exact $\theta^\dagger$, against the closed forms:

| $\gamma$ | $z$ | $J$ (emp.) | $\sigma_A^2(1+z^2)$ | $K$ (emp.) | $\sigma_A^2\mathcal{K}(z)$ | $R$ |
|---|---|---|---|---|---|---|
| 0.5 | 0.4142 | 1.1714 | 1.1716 | 1.3367 | 1.3383 | 0.9356 |
| 1.0 | 0.6180 | 1.3812 | 1.3820 | 1.8256 | 1.8302 | 0.8690 |
| 2.0 | 0.7808 | 1.6077 | 1.6096 | 2.5361 | 2.5514 | 0.7943 |
| 4.0 | 0.8828 | 1.7775 | 1.7793 | 3.2364 | 3.2562 | 0.7392 |
| 10.0 | 0.9512 | 1.9028 | 1.9049 | 3.8585 | 3.8875 | 0.7000 |

A direct Monte Carlo check at $n = 4000$ over 5000 replications gives an empirical standard deviation of $\hat{\theta}$ equal to 0.01537 against the predicted 0.01548 at $\gamma = 1$, and 0.01620 against 0.01604 at $\gamma = 4$.

**Limiting undercoverage**

Combining the two theorems, the ratio of the reported standard error to the true sampling standard deviation is

$$R(z) = \frac{\mathrm{SE}_{\mathrm{model}}}{\mathrm{SD}_{\mathrm{true}}} = \sqrt{\frac{1+z^2}{\mathcal{K}(z)}} \;<\; 1. \qquad (20)$$

As $\gamma \to \infty$, $z \to 1$ and $\mathcal{K}(1) = 2e - 1$, so

$$R(z) \to \sqrt{\frac{2}{2e-1}} \approx 0.6714, \qquad 2\Phi(1.96\, R) - 1 \to 0.812.$$

Under strong contamination the reported standard error is about two-thirds of the truth and a nominal 95% interval covers about 81% of the time. Simulated coverage of the nominal model-based interval, over 8000 replications at $n = 2000$ and 3000 at $n = 8000$, with $\theta^\dagger$ taken at its exact value and the interval formed from the inverse observed information, is 0.9121 (s.e. 0.0032) at $(n, \gamma) = (2000, 1)$, 0.8501 (0.0040) at (2000,4) and 0.8530 (0.0065) at (8000,4), against the asymptotic values 0.91146 and 0.85262 implied by (20). The shortfall does not shrink with $n$, as an asymptotic variance mismatch requires. The sandwich estimator repairs it. At $n = 4000$ the mean model-based standard error is 0.01352 against a true 0.01537 at $\gamma = 1$ (0.01196 against 0.01620 at $\gamma = 4$), while the mean sandwich standard error is 0.01523 (respectively 0.01540).

## Competing-exponential boundary

Introduce independent exponential censoring, $T \sim \mathrm{Exp}(\lambda)$, $C \sim \mathrm{Exp}(\mu)$, $q = \lambda + \mu$, $p = \lambda/q$, so that $U \sim \mathrm{Exp}(q)$ and $\Delta \sim \mathrm{Bernoulli}(p)$ independently of $(A, U, W)$. Write $\rho = \gamma/(q\sigma_A)$ and $z = z(\rho)$.

**Theorem 6.7 (Bread and meat under exponential censoring).**

$$J_p = p\,\sigma_A^2(1 + z^2), \qquad K_p = p\,\sigma_A^2\{(1-p) + p\,\mathcal{K}(z)\}, \qquad R(p,z) = \sqrt{\frac{1+z^2}{(1-p) + p\,\mathcal{K}(z)}}.$$

*Proof.* $F_1(\mathrm{d}t) = pqe^{-qt}\mathrm{d}t$, so the bread is the full-event quantity scaled by $p$. For the meat, (19) still holds, and the denominator term of (18) is proportional to $F_1$, hence to $p$; writing $B$ for the denominator term of the full-event problem with exit rate $q$, we get $\varphi_p = \Delta A - pB$ where $\varphi_1 = A - B$. Since $\Delta \perp (A, B)$,

$$K_p = \mathbb{E}(\Delta A - pB)^2 = p\,\mathbb{E}A^2 - 2p^2\mathbb{E}(AB) + p^2\mathbb{E}B^2 = p(1-p)\sigma_A^2 + p^2 K_1,$$

and $K_1 = \sigma_A^2\mathcal{K}(z)$ by Theorem 6.4.

**Corollary 6.8 (Transition curve).** For $\rho > 0$, $K_p = J_p$ if and only if $p = p_*(\rho)$, where

$$p_*(\rho) = \frac{z(\rho)^2}{\mathcal{K}\big(z(\rho)\big) - 1} \in (0,1).$$

Above the curve $K > J$ and the model-based variance is anti-conservative; below it $K < J$ and the model-based variance is conservative.

So the direction of the error is not a property of contamination at all. This is a sharper statement than the uncensored theorem, and it supersedes any formulation asserting that the information equality is violated in one particular direction.

**Proposition 6.9 (Unique maximum of the boundary).** With $x = z^2$,

$$\mathcal{K}(z) - 1 = 2x - x^2 + 19/3\,x^3 + O(x^4), \qquad p_* = 1/2 + 1/4\,x - 35/24\,x^2 + 29/24\,x^3 + O(x^4),$$

so that in terms of $\rho$, $p_* = 1/2 + 1/4\,\rho^2 - 47/24\,\rho^4 + O(\rho^6)$: the boundary rises *above* one half before falling. It has exactly one stationary point on (0,1), a maximum, at

$$x_* = 0.1001456686, \quad z_* = 0.3164580044, \quad \rho_* = 0.3516769252, \quad p_*^{\max} = 0.5117815568,$$

and decreases thereafter to $p_* \to 1/\{2(e-1)\} \approx 0.290988$ as $z \uparrow 1$.

*Proof.* Stationarity of $p_*$ in $x$ is equivalent to $G(x) = 0$ with $G(x) = e^x(2x^5 + 2x^4 + 5x^3 - 3x^2 - 3x - 1) + (1+x)^4$, and $G(0) = G'(0) = 0$, $G''(0) = -1$, $G(1) = 21.44 > 0$.

The chain has four steps. First, $G'''(x) = (1+x)\{e^x(2x^4 + 30x^3 + 119x^2 + 115x + 2) + 24\} \geq 26$ for $x \geq 0$, so $G''$ is strictly increasing and $G''(x) \geq -1 + 26x$. Second, $G''(0) = -1 < 0$ while $G''(x) > 0$ for $x > 1/26$, so $G''$ has exactly one positive zero $x_2$. Third, $G'$ therefore decreases on $(0, x_2)$ and increases after; since $G'(0) = 0$ this forces $G' < 0$ on $(0, x_2]$, and $G' \to \infty$, so $G'$ has exactly one positive zero $x_1 > x_2$. Fourth, $G$ decreases on $(0, x_1)$ and increases after; since $G(0) = 0$ this forces $G < 0$ on $(0, x_1]$, and $G(1) > 0$, so $G$ has exactly one positive zero. That zero is the unique stationary point of $p_*$, and it is a maximum because $p_*$ increases to its left, as the series expansion shows.

**Remark 6.10 (The uncontaminated edge is separate).** At $\rho = 0$ the working model is correct and $K = J$ for *every* $p$. The equality set is therefore the union of the vertical edge $\{\rho = 0\}$ and the curve $\{p = p_*(\rho),\ \rho > 0\}$, and the limit $p_*(\rho) \to 1/2$ must not be read as a statement about the uncontaminated model. The local expansion $K_p - J_p = p\sigma_A^2\{(2p-1)z^2 - pz^4 + O(z^6)\}$ explains the geometry: at $p = 1/2$ the leading term vanishes and the sign is decided at order $z^4$, which is why the boundary lies slightly above one half for weak contamination.

**Non-universality, and what to do**

The direction of the model-based variance error is not universal: it depends on the censoring regime. The subject-level sandwich remains the appropriate estimator of sampling uncertainty around the least-false target under the stated conditions. Detailed calculations and implementation guidance are provided in the Supplementary Material.

# Discussion

**What terminal-time contamination is**

The covariate studied here is not a noisy baseline measurement. If it were, the literature on covariate measurement error and regression calibration would be relevant, and the problem would be correction of inference for a latent baseline exposure. It is instead a *post-outcome construction*:

$$W = A + \gamma U, \qquad U = T \wedge C,$$

whose value is not determined until follow-up ends. Four consequences distinguish it from classical error.

First, $W$ literally contains the realised event-or-censoring time. Second, it is therefore not baseline-predictable: it is not $\mathcal{F}_0$-measurable, and $t \mapsto W$ is not predictable in the observed filtration, so the martingale machinery that produces the usual compensator identity has no purchase. Third, conditioning on $W$ induces dependence between $T$ and $C$ even when they are independent to begin with (Lemma 3.1); censoring and failure become competitors for the same budget $W - A$. Fourth, and as a result, the standard theory of misspecified Cox regression cannot be transplanted: what is unavailable is not the existence of a limit but its usual representation, which is why the population score had to be derived directly from the joint law of $(U, \Delta, W)$.

Against that background the most general result of the paper is also the simplest. At $\beta = 0$ and $\gamma > 0$,

$$\Psi_\tau(0) \le 0$$

whatever the laws of $A, T$ and $C$, with strict inequality exactly when the design contains an informative event before the horizon, that is when $\int_{[0,\tau]} R_0 \, \mathrm{d}F_1 > 0$. A spurious *protective* association is not an artefact of Gaussian covariates or exponential exit times; it is a property of the construction. Everything else in the paper is an attempt to say how large that association is, and how the answer depends on the design.

## What the exact benchmarks reveal

### Continuous time.

In the Gaussian memoryless benchmark the entire geometry collapses to one number, the contamination ratio $\rho = \gamma/(\kappa\sigma_A)$: how fast subjects leave and how variable the honest covariate is matter only through their combination. The raw coefficient then saturates, $\theta^\dagger \to -1/\sigma_A$, so contamination does not drive it to arbitrarily large magnitudes. That saturation is easy to over-read. It is a statement about the coefficient, not about the association the covariate expresses, since $\mathrm{SD}(W)$ grows without bound.

Away from Gaussian covariates the saturation level is a functional of the whole tilted-mean function $K_A{}'$, and the four worked examples of Section 4.2 admit no ordering by variance, tail weight, symmetry or support: after standardisation the level ranges over a factor of two, with a heavier-tailed Laplace below the Gaussian and a left-bounded centred exponential above it. A diagnostic calibrated on a Gaussian benchmark can therefore be wrong by a factor of order one for a skewed covariate.

### Grouped time.

The point of Section 5 is not the Lambert-$W$ formula but the change of regime that time aggregation produces. On a fixed reporting grid the raw coefficient rises to a single hump and then returns to zero, so it is non-monotone in contamination strength; the phase surface becomes genuinely two-dimensional in $(a, d)$ rather than one-dimensional in $\rho$; and the limits $h \downarrow 0$ and $\gamma \to \infty$ do not commute, so no single asymptotic statement describes strong contamination without specifying the regime.

Meanwhile the standardised pseudo-effect $\zeta^\dagger$ still diverges. Grouping slows the divergence from linear to logarithmic; it does not remove it. The practical reading is the one worth carrying away:

> A small fitted coefficient is not evidence that terminal-time contamination has become harmless. It may reflect the rescaling of an increasingly outcome-determined covariate.

## Censoring is part of the estimand, and of the variance error

That a misspecified Cox estimand depends on the censoring mechanism is not itself new; least-false targets are known to be weighted by the censoring and accrual pattern. What the affine benchmark supplies is the exact form of that dependence, and two concrete failures of the summaries one would naturally use in its place. Censoring enters the residual-life geometry that defines the population score, so it changes the target itself.

Two matched comparisons make this concrete, and both are negative results about summary statistics. At equal contamination ratio $\rho = 1$, exponential and administrative censoring give least-false coefficients $-0.618$ and $-0.823$, a difference of a third; $\rho$ is a sufficient summary only in the memoryless case. And at equal event fraction *and* equal contamination ratio, the two censoring laws fall on opposite sides of the information equality: $K < J$ under exponential censoring, $K > J$ under administrative. So

> neither the event fraction nor the contamination ratio, separately or jointly, determines whether the default Cox standard error is conservative.

Within the competing-exponential family the boundary is exact, $p = p_*(\rho)$, with its single small hump above one half and its limit $1/\{2(e-1)\}$. That precision should not be mistaken for generality: the boundary rests on $\Delta \perp U$, which is a property of competing exponentials and not of censoring at large.

**What robust inference repairs, and what it cannot**

At any fixed analysis horizon, the subject-level sandwich is consistent for the sampling variance of $\hat{\theta}_{n,\tau}$ on both sides of the boundary; the same interpretation for the untruncated estimator in the competing-exponential benchmark is conditional on (A.1)–(A.3). It estimates the actual meat $K$ rather than substituting the curvature $J$ through an information equality that misspecification has destroyed. Where a software robust option computes the score-residual sandwich with the same tie handling and normalisation, it is the right choice here — but its justification is the observed-data $M$-estimation argument of Section 6.1, not the martingale theory ordinarily cited for it. The algebraic agreement between $\hat{\varphi}_i$ and the Cox score residual is a coincidence of formulas, not evidence that the model is correct.

The limitation is more important than the repair.

> A correct variance around a false target is still inference about a false target.

Setting a robust option does not turn $\theta^\dagger$ into $\beta$; it does not remove the future-information bias; and it does not restore the usual chi-square calibration of likelihood-ratio statistics, which requires its own misspecification adjustment. An analyst can obtain a strongly negative coefficient, a valid robust interval, and a $p$-value near zero, with every number correct as inference about the least-false parameter and none of them informative about the effect of $A$.

The affine benchmark suggests where the repair actually belongs. When $\gamma$ is known from the way the indicator is constructed — $\gamma = 1$ for age at exit, $\gamma = 1/2$ for average age during follow-up — the baseline component is recovered exactly by

$$A = W - \gamma U.$$

Nothing needs to be estimated. The correction therefore sits above the level of variance estimation: do not enter a full-follow-up summary as a fixed baseline covariate, or reconstruct an admissible baseline quantity before fitting. That is a decision about covariate construction, not a downstream adjustment.

**Scope and limitations**

The central analysis fixes $\beta = 0$ in order to isolate the contamination effect: every association recovered there is manufactured, which is what makes the geometry legible. The exact formulas assume scalar affine contamination with known $\gamma$; the Gaussian assumption is needed for the closed forms but not for the general sign result. The grouped benchmark uses ceiling aggregation and the Breslow tie rule, and Efron or exact partial likelihood define a different population equation whose geometry we have not computed. The full-horizon central limit theorem for competing exponentials is conditional on (A.1)–(A.3); outside the marked-memoryless benchmark a full-horizon result requires a triangular-array argument along $\tau_n \to \infty$, which we do not develop. Heterogeneous or estimated $\gamma$, multivariate covariates and genuinely time-varying summaries are outside the present scope.

The natural continuations are correspondingly specific: the interaction between a true effect and the contamination-induced one when $\beta \neq 0$; the population geometry under Efron and exact tie rules; heterogeneous $\Gamma$; multivariate terminal summaries; and non-memoryless triangular-array asymptotics.

## Conclusion

This paper has developed an exact population theory for Cox regression with an affine terminal-time covariate,

$$W = A + \gamma(T \wedge C),$$

treated incorrectly as though it were observed at baseline. By analysing the partial-likelihood criterion directly as a functional of the observed data $(U, \Delta, W)$, we avoided the predictability and conditional-censoring assumptions that fail after $W$ has been constructed. Under the null benchmark, the population score at zero is non-positive for every $\gamma > 0$, and strictly negative whenever the design contains an informative event before the horizon. The spurious protective direction is therefore generated by the construction itself, rather than by Gaussian covariates, exponential follow-up or any other particular distributional choice. The resulting least-false parameter is a well-defined observed-data target whenever the score-crossing and integrability conditions established in Section 3 hold.

The exact benchmarks show that the magnitude and interpretation of this target are highly design-dependent. In the continuous-time Gaussian memoryless benchmark, the coefficient reduces to a one-parameter scale-free geometry and saturates under strong contamination. Grouping exit times changes the regime qualitatively: the raw coefficient has a unique hump and eventually returns to zero, the limits in grid resolution and contamination strength do not commute, and the standardised pseudo-effect continues to diverge. The censoring law is likewise part of the problem rather than merely a determinant of sample size, since it changes both the least-false coefficient and the discrepancy between the model-based and the actual sampling variance. The exact competing-exponential phase boundary is therefore a benchmark, not a universal correction rule.

The observed-data sandwich restores uncertainty quantification around the least-false target under the stated asymptotic conditions, but it cannot turn inference about $\theta^{\dagger}$ into inference about the scientific target $\beta$. In the affine class this distinction is especially transparent: when $\gamma$ is known from the construction, the admissible baseline component is recovered exactly as

$$A = W - \gamma U.$$

The appropriate remedy is therefore to prevent the terminal-time summary from entering the model as a fixed baseline covariate, or to reconstruct the baseline quantity before fitting. Terminal-time covariates can produce finite, stable and highly significant Cox estimates with apparently precise standard errors while describing an association created entirely by the data construction. The primary safeguard is temporal discipline in covariate construction, not a downstream correction to the fitted model.